\documentclass{article}

\usepackage{arxiv}

\usepackage[utf8]{inputenc} 
\usepackage[T1]{fontenc}    
\usepackage[colorlinks, citecolor=green, linkcolor=red]{hyperref}      
\usepackage{url}            
\usepackage{booktabs}       
\usepackage{amsfonts}       
\usepackage{nicefrac}       
\usepackage{microtype}      
\usepackage{lipsum}		
\usepackage{graphicx}
\usepackage{doi}
\usepackage{authblk}
\usepackage{framed,multirow}

\usepackage{amssymb}
\usepackage{latexsym}
\usepackage{graphicx}
\usepackage{amsmath}
\usepackage{amssymb}
\usepackage{booktabs} 
\usepackage{tabularx}
\usepackage{array}
\usepackage{booktabs} 
\usepackage{tabularx, multirow}
\usepackage{tabularx, multirow, booktabs}
\usepackage{subcaption}
\usepackage{algorithm}
\usepackage{authblk}
\usepackage{float}
\usepackage[nameinlink, noabbrev]{cleveref} 
\usepackage{graphicx}%
\usepackage{multirow}%
\usepackage{amsmath,amssymb,amsfonts}%
\usepackage{amsthm}%
\usepackage{mathrsfs}%
\usepackage[title]{appendix}%
\usepackage{xcolor}%
\usepackage{textcomp}%
\usepackage{manyfoot}%
\usepackage{booktabs}%
\usepackage{algorithm}%
\usepackage{algorithmicx}%
\usepackage{algpseudocode}%
\usepackage{listings}%
\usepackage{url}          
\usepackage{nicefrac}       
\usepackage{microtype}      
\usepackage{lipsum}		
\usepackage{doi}
\usepackage{framed,multirow}

\usepackage{amssymb}
\usepackage{latexsym}
\usepackage{graphicx}
\usepackage{amsmath}
\usepackage{amssymb}
\usepackage{booktabs} 
\usepackage{tabularx}
\usepackage{array}
\usepackage{booktabs} 
\usepackage{tabularx, multirow}
\usepackage{tabularx, multirow, booktabs}
\usepackage{float}
\usepackage{amsmath}
\usepackage{multirow}
\usepackage{colortbl}
\usepackage{makecell}
\usepackage{subcaption}
\usepackage{array}
\usepackage[nameinlink, noabbrev]{cleveref} 

\theoremstyle{thmstyleone}%
\theoremstyle{thmstyletwo}%

\theoremstyle{thmstylethree}%
\crefformat{figure}{#2{\color{red}Figure~#1}#3}
\crefformat{section}{#2{\color{red}Section~#1}#3}
\crefformat{equation}{#2{\color{red}Equation~#1}#3}
\crefformat{table}{#2{\color{red}Table~#1}#3}
\crefformat{algorithm}{#2{\color{red}Algorithm~#1}#3}

\usepackage{url}
\usepackage{xcolor}
\newcolumntype{C}[1]{>{\centering\arraybackslash}p{#1}}

\newcolumntype{H}{>{\centering\arraybackslash}X}

\usepackage{enumitem}
\newlist{boldenum}{enumerate}{1}
\setlist[boldenum]{label=\textbf{(\arabic*)}}

\definecolor{newcolor}{rgb}{.8,.349,.1}

\title{\Large Bridging Classical and Quantum Machine Learning: Knowledge Transfer from Classical to Quantum Neural Networks Using Knowledge Distillation}

\author[1]{\large Mohammad Junayed Hasan}
\author[2, *]{\large M.R.C. Mahdy}

\affil[1]{Department of Computer Science, Johns Hopkins University, Baltimore, MD, USA}

\affil[2]{Department of Electrical and Computer Engineering, North South University, Bashundhara, Dhaka}

\hypersetup{
pdftitle={A template for the arxiv style},
pdfsubject={q-bio.NC, q-bio.QM},
pdfauthor={David S.~Hippocampus, Elias D.~Striatum},
pdfkeywords={First keyword, Second keyword, More},
}

\begin{document}

\maketitle

\begin{abstract}
    Quantum neural networks (QNNs), harnessing superposition and entanglement, have shown potential to surpass classical methods in complex learning tasks but remain limited by hardware constraints and noisy conditions. In this work, we present a novel framework for transferring knowledge from classical convolutional neural networks (CNNs) to QNNs via knowledge distillation, thereby reducing the need for resource-intensive quantum training and error mitigation. We conduct extensive experiments using two parameterized quantum circuits (PQCs) with 4 and 8 qubits on MNIST, FashionMNIST, and CIFAR10 datasets. The approach demonstrates consistent accuracy improvements attributed to distilled knowledge from larger classical networks. Through ablation studies, we systematically compare the effect of state-of-the-art dimensionality reduction techniques—fully connected layers, center cropping, principal component analysis, and pooling—to compress high-dimensional image data prior to quantum encoding. Our findings reveal that fully connected layers retain the most salient features for QNN inference, thereby surpassing other downsampling approaches. Additionally, we examine state-of-the-art data encoding methods (amplitude, angle, and qubit encoding) and identify amplitude encoding as the optimal strategy, yielding superior accuracy across all tested datasets and qubit configurations. Through computational analyses, we show that our distilled 4-qubit and 8-qubit QNNs achieve competitive performance while utilizing significantly fewer parameters than their classical counterparts. These outcomes highlight the potential of quantum-classical knowledge exchange in enhancing QNN scalability and efficiency. Our results establish a promising paradigm for bridging classical deep learning and emerging quantum computing, paving the way for more powerful, resource-conscious models in quantum machine intelligence.
\end{abstract}


\section{Introduction}

Quantum machine learning (QML) exhibits remarkable potential to transform computational paradigms by leveraging the principles of quantum superposition and entanglement. This can, in specific scenarios, lead to exponential speedups and richer model expressivity than classical counterparts. In particular, quantum neural networks (QNNs) have recently demonstrated compelling advantages, including faster convergence and enhanced classification accuracy, even when constrained to a similar number of trainable parameters as classical networks \cite{potempa2022comparing, alam2022qnet}. Moreover, recent theoretical insights confirm that the effective dimension of well-designed QNNs can exceed that of classical models, hinting at a superior capacity to tackle intricate data distributions \cite{coles2021seeking}.

Nonetheless, the nascent status of QML introduces intrinsic obstacles. Modern quantum hardware is restricted by factors such as limited qubit availability, short coherence times, and susceptibility to noise and decoherence \cite{beer2020training, du2021learnability}. These challenges severely constrain the scale and complexity of QNNs that can be reliably executed, further exacerbated by stochastic outcomes inherent to quantum sampling \cite{cerezo2022challenges}. As a result, current quantum platforms often demand extensive resources to mitigate errors, leading to a steep computational overhead when attempting to train deeper, more accurate QNNs \cite{kwak2021quantum, zhao2021review}. A variety of strategies have been devised to alleviate errors and resource limitations in QNNs, including variational error-mitigation schemes, resource-efficient quantum circuit optimizations, and quantum-to-quantum or classical-to-quantum transfer learning methods. Yet, these approaches remain works in progress and face distinct limitations. For instance, neural error mitigation can require large labeled datasets and computational power to train a secondary neural corrector \cite{alam2023knowledge}, while hybrid variational error mitigation introduces additional parameters that may escalate optimization complexities. Circuit simplifications and feature-map optimizations \cite{shin2020knowledge} are often hardware- or application-specific, restricting their broader applicability. Quantum-to-quantum transfer learning presupposes available large-scale quantum models to serve as teachers, whereas classical-to-quantum transfer still primarily depends on classical representations that may not capture complex quantum correlations or entanglement.
\begin{figure*}
    \centering
    \includegraphics[width=1\linewidth]{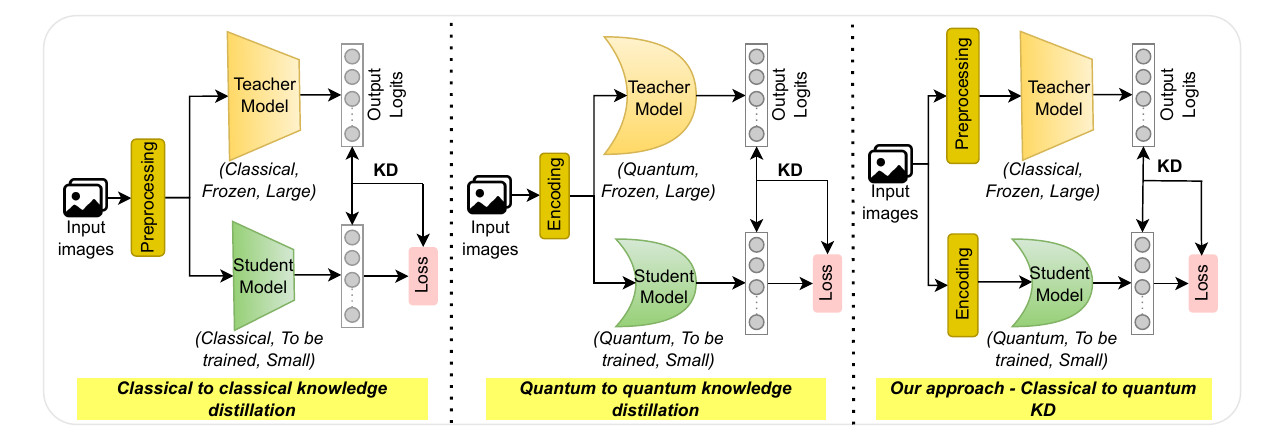}
    \caption{Existing approaches of classical to classical and quantum to quantum knowledge distillation, compared to our proposed method: fusion of classical and quantum approaches through classical to quantum knowledge distillation}
    \label{fig:intro}
\end{figure*}
Knowledge distillation \cite{hinton2015distilling} provides a powerful mechanism to sidestep these issues by allowing a large, potentially resource-intensive ``teacher'' model to impart its learned representations to a smaller ``student'' model. This paradigm obviates the need for extensive quantum training, as the distilled student network primarily relies on the teacher’s logits to guide its learning. Crucially, distillation can mitigate noise sensitivity, diminish reliance on cumbersome quantum resources, and unveil deeper structural insights in the student network. Consequently, knowledge distillation is emerging as a robust technique for designing and optimizing neural architectures.

While a handful of studies have examined quantum-to-quantum transfer learning or response-based knowledge distillation in QNNs \cite{alam2023knowledge}, the domain of classical-to-quantum knowledge transfer remains underexplored. Considering the inherent constraints of QML, it is vital to investigate how pre-trained classical models can effectively enhance QNNs through a mere transfer of their outputs. This study thus addresses the following research questions:
\begin{itemize}
    \item \textbf{Research Question 1:} Can knowledge be transferred from a classical neural network (teacher) to a quantum neural network (student) through knowledge distillation?
    \item \textbf{Research Question 2:} How does this distillation from classical networks impact the performance of a QNN in real-world image classification tasks?
    \item \textbf{Research Question 3:} Can such distillation strategies be leveraged to significantly improve resource-constrained quantum models, thereby minimizing the need for large-scale quantum training?
\end{itemize}

To address these questions, we propose a novel knowledge distillation framework that transfers learned representations from multiple classical teacher models to smaller QNN student models. As illustrated in Fig.~\ref{fig:intro}, this approach bridges the classical and quantum paradigms by utilizing the logits of high-performing convolutional neural networks (CNNs)---LeNet, AlexNet, ResNet, DenseNet, and EfficientNet---to train 4-qubit and 8-qubit variational QNNs. Our experiments span three popular benchmarks: MNIST \cite{deng2012mnist}, FashionMNIST \cite{xiao2017fashion}, and CIFAR10 \cite{krizhevsky2009learning}. We compare quantum student models with and without distillation and further evaluate their performances through thorough statistical significance testing, ablation studies on dimensionality reduction and data encoding strategies, and a comprehensive error analysis. Our principal contributions can be summarized as follows:
\begin{itemize}
    \item \textbf{Hybrid Distillation Framework:} We introduce a robust knowledge distillation pipeline that enables the reuse of existing pre-trained classical models for training resource-efficient QNNs, effectively eliminating the need for extensive quantum-specific error mitigation.
    \item \textbf{Systematic Performance Evaluation:} We quantify the gains from classical-to-quantum distillation across three datasets and multiple state-of-the-art teacher models, showcasing significant improvements in QNN accuracy.
    \item \textbf{Comparison with State-of-the-art Techniques:} Through controlled ablation experiments on dimensionality reduction and data encoding, we identify best practices for efficiently embedding high-dimensional data into few-qubit circuits. Further, our findings are validated by rigorous statistical tests.
    \item \textbf{Generalizability and Future Outlook:} We discuss broader implications of the classical-to-quantum transfer paradigm, highlighting how it can empower cross-domain applications and pave the way for more robust quantum-classical hybrid intelligence.
\end{itemize}

\section{Related Works}
\label{sec:related_work}
\subsection{Problems and Challenges in QML}
QML aspires to harness the properties of quantum mechanics to revolutionize computation, but the path is restricted with various challenges and obstacles. The concept of quantum supremacy, where quantum systems perform tasks beyond the reach of classical computers, has seen experimental strides as evidenced by \cite{arute2019quantum} and \cite{madsen2022quantum}. However, these pioneering steps also underscore the fragility of quantum states and their vulnerability to decoherence, a problem that Shor’s early work \cite{shor1995scheme} sought to address through quantum error correction, a technique still pivotal in contemporary research \cite{bharti2022noisy}. The transient stability of qubits, a topic explored by \cite{zwerver2022qubits}, and the intricate challenge of manipulating quantum states, as demonstrated by \cite{okamoto2013coherent}, add layers of complexity to quantum computing.

The Noisy Intermediate-Scale Quantum (NISQ) era, defined by the presence of noisy quantum bits, has constrained the scalability of quantum computers, a concern that  \cite{amin2018quantum} aimed to tackle through the quantum Boltzmann machine, an approach designed to work within these noisy environments. Concurrently, the development of quantum algorithms and neural networks that can operate under such noise is the focus of \cite{biamonte2017quantum} and \cite{abbas2021power}, who seek methods to optimize QML models despite these limitations. Data encoding poses a unique challenge in QML, as most algorithms are tailored for inherently quantum data. \cite{havlivcek2019supervised} delve into creating quantum-enhanced feature spaces, while \cite{saggio2021experimental} look into reinforcement learning agents that can operate within the quantum realm. These studies offer insights into the potential application domains for QML, suggesting that despite current limitations, certain tasks may still be within reach. Furthermore, the practical implementation of quantum computing and the exploration of its utility across various domains have been the subject of investigation by \cite{cerezo2021variational}, who discuss the variational quantum algorithms that could be the cornerstone of future quantum applications.

Despite the under-developed stage of QML, there is an ongoing effort to address its challenges. \cite{bouland2020prospects} provides a comprehensive analysis of the prospects and barriers that lie ahead, ensuring that the community remains grounded in reality while exploring the vast potential of quantum technologies. The integration of efforts from all these studies manifests a commitment to overcoming the current hurdles, indicating a future where quantum and classical computing may converge to solve problems that were once thought unsolvable. In this research, we commit to this challenge and address it using a hybrid classical-quantum architecture and demonstrate that the methodology takes handling the challenges faced in the realm of QML one step further.

\subsection{Error Mitigation and Resource Handling}
The pursuit of reliable quantum computing is marked by a need to address the fundamental limits of quantum error mitigation. Works such as those by \cite{takagi2022fundamental} provide a foundational understanding of the theoretical bounds in quantum systems, crucial for error correction techniques in neural networks. This is further explored through the lens of scalability and statistics of errors in recent advancements within noisy intermediate-scale quantum (NISQ) technologies \cite{qin2023error}. Reference-state error mitigation strategies \cite{lolur2023reference} have shown promise in improving the accuracy of quantum computations, which is vital for the fidelity of knowledge distillation processes. Moreover, the extension of computational reach through error mitigation in noisy quantum processors \cite{kandala2019error} has direct implications for the robust execution of QNNs. The integration of learning-based techniques for quantum error mitigation \cite{strikis2021learning} parallels the adaptive nature of knowledge distillation, suggesting a methodological symmetry that could be harnessed for transferring knowledge. The evaluation and benchmarking of error mitigation techniques are critical for determining the most effective methods for application in QNNs \cite{bultrini2023unifying}. This selection is further informed by the development of a unified approach to data-driven error mitigation \cite{lowe2021unified}, offering a pathway to harnessing empirical data in the error correction process. Techniques for mitigating errors in quantum approximate optimization \cite{weidinger2023error} can also inform the optimization strategies within the knowledge distillation framework.

On the front of resource handling, the importance of optimizing resource efficiencies is underscored by \cite{fellous2023optimizing}, drawing attention to the balance between computational capabilities and resource constraints. This optimization is directly applicable to the development of QNNs, where the allocation and management of quantum resources are crucial. Furthermore, the critical challenges of scaling quantum computation, as outlined by  \cite{martonosi2019next}, must be addressed to enable the practical implementation of fault-tolerant QNNs. Insights into the cross-discipline theoretical research on quantum computing, such as those presented by \cite{ayoade2022artificial}, highlight the urgency and significance of resource management in the context of knowledge distillation. Additionally, the potential applications of quantum computing in specific domains like renewable energy optimization \cite{giani2021quantum} illustrate the diverse utility and the resource handling strategies that may be required for energy-efficient QNN operations. Differently from previous approaches, we completely nullify the need for QNN optimization using error mitigation or resource handling. Rather, keeping the student quantum model small and static, we depend fully on pre-trained classical models to improve the performance of the QNN, thus addressing the challenges of QML in a new, out-of-the-box method.

\subsection {Transfer Learning and Knowledge Transfer in QML}
Transfer learning has emerged as a potent strategy in classical machine learning, enabling models trained on one task to be repurposed for another related task \cite{zen2020transfer,smith2018outsmarting}. In QML, this approach promises to address scalability and adaptability challenges that are crucial in the practical application of neural-network quantum states (NNQS) \cite{zen2020transfer}. Scalability, a primary concern in QML, is particularly relevant for tasks that require models to extrapolate learned behaviors to larger quantum systems. Protocols inspired by physics have shown that features learned from smaller systems can significantly accelerate and refine the learning process when applied to more complex systems, as evidenced in the one-dimensional transverse field Ising and Heisenberg XXZ models \cite{zen2020transfer}. The versatility of transfer learning is further showcased by its application to the optimization of neural network potentials, exemplified by ANI-1x, which was retrained to near gold-standard quantum mechanical accuracy. This highlights transfer learning's broad potential across disciplines, including materials science, biology, and chemistry \cite{smith2018outsmarting,smith2019approaching}. 

Integrating classical and quantum neural networks has extended transfer learning protocols into classical-quantum hybrid domains. By combining pre-trained classical networks with variational quantum circuits, researchers have opened up new possibilities for processing high-dimensional data such as images before entangling them within the quantum processing framework \cite{mari2020transfer}. This hybrid approach not only utilizes the strengths of classical networks in data handling but also harnesses the quantum advantages for feature processing, potentially revolutionizing fields like image recognition and quantum state classification \cite{mari2020transfer,umer2022integrated}. Moreover, the concept of knowledge distillation has been introduced into the realm of QML, allowing the creation of new QNNs through approximate synthesis \cite{alam2023knowledge}. This methodology enables the reduction of circuit layers or the alteration of gate sets without the necessity of training from scratch. Empirical analyses suggest that significant reductions in circuit complexity can be achieved while simultaneously improving accuracy, even under noisy conditions, underscoring the effectiveness of transfer learning techniques in the development of QNNs \cite{alam2023knowledge}. More recently, hybrid-classical quantum convolutional neural networks have seen applications in brain tumor classification \cite{choudhuri2023brain}, brain lesion detection \cite{amin2023detection} and pneumonia detection \cite{kulkarni2023classical}.

In terms of practical applications, transfer learning has been successfully applied to improve the accuracy and efficiency of surrogate models used in design optimization \cite{kaya2019using}, and disease detection \cite{amin2022new,tamilvizhi2022quantum}, demonstrating its efficacy in enhancing computational time and model performance. Furthermore, leveraging pre-existing computational datasets through deep transfer learning constructs robust prediction models that outperform those based solely on theoretical calculations \cite{jha2019enhancing}. The domain of spoken command recognition (SCR) has also benefitted from the implementation of hybrid transfer learning algorithms. By transferring knowledge from a pre-trained classical network to a hybrid quantum-classical model, researchers have been able to significantly improve performance on SCR tasks, showcasing the practical benefits of transfer learning in QML \cite{qi2022classical}.

Quantum Convolutional Neural Networks (QCNNs) are gaining popularity in both quantum and classical data classification recently \cite{kim2023classical}. The adoption of transfer learning in QML is proposed to circumvent the scalability constraints of quantum circuits in the noisy intermediate-scale quantum (NISQ) era. For instance, research has shown that small QCNNs can effectively solve complex classification tasks by leveraging pre-trained classical CNNs, thus bypassing the need for expansive quantum circuitry \cite{qi2022classical}. This approach is validated through numerical simulations of QCNNs, which demonstrate a superior classification accuracy over classical models when pre-trained on different datasets. The synergy between classical and quantum networks is evident when a classical CNN, trained on FashionMNIST data, is utilized to enhance the performance of a QCNN for MNIST data classification \cite{kim2023classical}. Moreover, very recent studies have also calibrated large language models in QNNs \cite{neto2024quantum} and accomplished hybrid classical-quantum transfer learning for text classification \cite{ardeshir2024hybrid}. However, differently from \cite{qi2022classical} and \cite{kim2023classical}, we use knowledge distillation and show empirically that knowledge transfer is possible across homogeneous datasets in quantum models by simply trying to mimic the soft targets of classical CNNs instead of hectic training by freezing the pre-trained model and transferring the knowledge to a QNN during training in the previous approaches.

\section{Preliminaries}
\subsection{Quantum Superposition and the Mathematical Intuition Behind QNNs}
A QNN can be represented by a parameterized quantum circuit (PQC) \( U(\boldsymbol{\theta}) \), where \( \boldsymbol{\theta} \) denotes the vector of parameters. The output of the QNN for an input state \( |\psi(\mathbf{x})\rangle \) prepared by a quantum feature map is given in Eq. \ref{eq:eq5} by:
\begin{equation}
\label{eq:eq5}
f_s(\mathbf{x}) = \langle \psi(\mathbf{x}) | U^\dagger(\boldsymbol{\theta}) O U(\boldsymbol{\theta}) | \psi(\mathbf{x}) \rangle,
\end{equation}
where \( O \) is an observable corresponding to the output measurement.

The postulates of quantum mechanics \cite{von2018mathematical} allow us to describe the state of a quantum system using a wave function or state vector in a Hilbert space. A quantum system in a state \( |\psi\rangle \) can exist in a superposition of different basis states \( \{|i\rangle\} \). This superposition is mathematically represented in Eq. \ref{eq:eq6} as:

\begin{equation}
\label{eq:eq6}
|\psi\rangle = \sum_i c_i |i\rangle,
\end{equation}
where \( c_i \) are complex coefficients satisfying the normalization condition \( \sum_i |c_i|^2 = 1 \).

The probability \( P(i) \) of the system being found in a particular basis state \( |i\rangle \) upon measurement is given by the square of the modulus of the coefficient corresponding to that state: $P(i) = |c_i|^2$

The expectation value \( \langle O \rangle \) of an observable \( O \), which corresponds to a physical quantity that can be measured, is computed according to Eq. \ref{eq:eq7} as:

\begin{equation}
\label{eq:eq7}
\langle O \rangle = \langle \psi | O | \psi \rangle = \sum_{i,j} c_i^* c_j \langle i | O | j \rangle.
\end{equation}

For an observable \( O \) with eigenstates \( |o_k\rangle \) and eigenvalues \( o_k \), the expectation value takes the form of Eq. \ref{eq:eq8}:

\begin{equation}
\label{eq:eq8}
\langle O \rangle = \sum_k o_k P(o_k),
\end{equation}
where \( P(o_k) \) is the probability of obtaining the eigenvalue \( o_k \) upon measuring the observable \( O \).

In the context of quantum computing, particularly in quantum neural networks, the output of the network can be associated with the expectation value of a certain observable. By designing the observable suitably, the expectation values can be interpreted as probabilities, providing us with a powerful tool to map quantum computations to classical outputs useful for tasks such as classification. This probabilistic interpretation is the key to the integration of quantum systems with machine learning, as it allows us to use quantum computations to produce outputs that can be understood and utilized within the classical framework of neural networks.

The knowledge transfer from the classical teacher model to the quantum student model is achieved by the optimization of the quantum circuit parameters \( \boldsymbol{\theta} \) to minimize \( \mathcal{O}_{total} \). This optimization can be performed using gradient-based methods where the gradient can be estimated via the parameter shift rule or other quantum backpropagation techniques.

\subsection{Data Encoding in QNNs}
\label{sec:encode_preli}
\subsubsection{Angle Encoding}

Angle encoding, also known as rotational encoding, is a method used in quantum computing to encode classical data into the parameters of quantum gates. This technique utilizes the rotation gates such as RX, RY, and RZ to map classical data points to angles of rotation, effectively embedding the data into the quantum state.

Given a classical vector \( \mathbf{x} \in \mathbb{R}^n \), each element \( x_i \) of the vector can be associated with a rotation angle for a corresponding quantum gate. The encoding can be represented as a sequence of operations on a quantum state \( |\psi\rangle \) in a Hilbert space of \( n \) qubits, as shown in Eq. \ref{eq:eq21}:

\begin{equation}
\label{eq:eq21}
|\psi(\mathbf{x})\rangle = \bigotimes_{i=0}^{n-1} R_{\phi_i}(x_i) |0\rangle,
\end{equation}
where \( R_{\phi_i} \) denotes the rotation gate applied to the \( i \)-th qubit with \( \phi_i \) being one of the axes \( \{X, Y, Z\} \), and \( x_i \) is the classical data point determining the rotation angle.

The angle encoding process involves applying a series of rotation gates to an initial quantum state, usually the zero state \( |0\rangle^{\otimes n} \), transforming it into a state that represents the encoded classical data. The choice of rotation axis and the order of gate application can vary depending on the specific implementation and the desired properties of the encoded state.

Angle encoding is particularly useful in variational quantum circuits and quantum machine learning models, where the parameters of the quantum gates are optimized during the training process. This form of encoding allows the model to directly manipulate the quantum state using classical data, enabling the exploration of complex feature spaces through the quantum system's inherent properties of superposition and entanglement.

\begin{equation}
\label{eq:eq22}
R_{\phi_i}(x_i) = e^{-i \frac{x_i}{2} \sigma_{\phi_i}},
\end{equation}
where \( \sigma_{\phi_i} \) is the Pauli operator corresponding to the rotation axis \( \phi_i \), and \( x_i \) is the angle of rotation derived from the classical data.

By employing angle encoding, a QNN can harness the full potential of quantum computation to process and learn from classical data, potentially leading to more powerful and efficient machine learning models.

\subsubsection{Amplitude Encoding}
Amplitude encoding is another technique used to embed classical data into the quantum state of a quantum system. Unlike angle encoding, which relies on rotation gates, amplitude encoding directly maps the classical data to the amplitudes of the quantum state's basis vectors.

Given a classical vector $\mathbf{x} = (x_0, x_1, \ldots, x_{n-1})$, the amplitude encoding can be represented as:

\begin{equation}
\label{eq:amp_encoding}
|\psi(\mathbf{x})\rangle = \sum_{i=0}^{2^n-1} x_i |i\rangle,
\end{equation}
where $|i\rangle$ represents the computational basis state corresponding to the binary representation of the index $i$, and the amplitudes $x_i$ are derived from the classical data $\mathbf{x}$.

The normalization condition for the quantum state must be satisfied:

\begin{equation}
\label{eq:amp_norm}
\sum_{i=0}^{2^n-1} |x_i|^2 = 1.
\end{equation}

Amplitude encoding allows for a direct mapping of classical data to the quantum state, potentially enabling more efficient processing and manipulation of the encoded information. However, it comes with the challenge of ensuring the normalization condition is met, which may require additional preprocessing or normalization steps.

\subsubsection{Qubit Encoding}
Qubit encoding, also known as basis encoding, represents classical data by associating each basis state of a quantum system with a specific data point. This encoding method is particularly suitable for discrete or categorical data.

Consider a classical dataset $\mathcal{D} = \{d_1, d_2, \ldots, d_N\}$ consisting of $N$ distinct data points. Each data point $d_i$ can be mapped to a computational basis state $|i\rangle$ of an $n$-qubit system, where $n = \lceil\log_2 N\rceil$.

The encoding process can be represented as:

\begin{equation}
\label{eq:qubit_encoding}
\mathcal{E}: \mathcal{D} \rightarrow \{|0\rangle, |1\rangle, \ldots, |2^n-1\rangle\},
\end{equation}
where $\mathcal{E}$ is the encoding function that maps each data point $d_i$ to the corresponding computational basis state $|i\rangle$.

To encode a specific data point $d_i$, the quantum state is prepared in the basis state $|i\rangle$ using a sequence of quantum gates, such as Hadamard gates and controlled-NOT gates.

Qubit encoding is particularly useful when dealing with discrete or categorical data, as it allows for a direct mapping of data points to the computational basis states. However, for continuous data, additional preprocessing or discretization steps may be required before encoding.

\subsection{QNN Training}

In the training of QNNs, the optimization of parameters \( \boldsymbol{\theta} \) is crucial. Due to the nature of quantum computing, traditional backpropagation as used in classical neural networks is not directly applicable. Instead, we utilize the parameter shift rule to compute gradients of quantum circuits.

\subsubsection{Parameter Shift Rule}
The parameter shift rule is a method to compute the gradient of an expectation value of an observable with respect to the parameters of a quantum circuit \cite{liu2018differentiable}. For a parameter \( \theta_j \) and an observable \( O \), the gradient can be calculated using the following Eq. \ref{eq:eq9} as:

\begin{equation}
\label{eq:eq9}
\frac{\partial \langle O \rangle}{\partial \theta_j} = \frac{1}{2}\left[\langle O \rangle_{\theta_j+\frac{\pi}{2}} - \langle O \rangle_{\theta_j-\frac{\pi}{2}}\right],
\end{equation}
where \( \langle O \rangle_{\theta_j \pm \frac{\pi}{2}} \) denotes the expectation value of \( O \) when the parameter \( \theta_j \) is shifted by \(\pm \frac{\pi}{2}\) respectively.

\subsubsection{Quantum Gradient Descent}
Once the gradients are computed, the parameters can be updated using a gradient descent algorithm. The update rule at each iteration \( t \) is given by Eq. \ref{eq:eq10}:

\begin{equation}
\label{eq:eq10}
\theta_j^{(t+1)} = \theta_j^{(t)} - \eta \frac{\partial \mathcal{O}_{total}}{\partial \theta_j},
\end{equation}
where \( \eta \) is the learning rate.

The learning rate can be fixed or adaptively changed according to specific strategies, such as learning rate annealing \cite{ward2020adagrad} or using advanced optimizers like Adam \cite{kingma2014adam}, RMSprop \cite{mukkamala2017variants}, or AdaGrad \cite{duchi2011adaptive}, which are adapted for the quantum domain.

\subsubsection{Quantum Circuit Learning}
The training process involves iteratively adjusting the parameters \( \boldsymbol{\theta} \) to minimize the loss function \( \mathcal{O}_{total} \). This is referred to as quantum circuit learning and is given by Eq. \ref{eq:eq4}. The optimization process leverages the quantum superposition and entanglement to explore the parameter space more efficiently than classical methods. The inherent probabilistic nature of quantum measurements is accounted for in the optimization loop, making the training resilient to the stochastic nature of quantum mechanics. By integrating these quantum-specific optimization techniques, we aim to exploit the full potential of QNNs while addressing the challenges posed by the hardware limitations and the unique aspects of quantum computation.

\subsubsection{Quantum Measurement and Quantum to Classical Decoding}
After the optimization of the QNN's parameters $\boldsymbol{\theta}$, the final step is to measure the quantum state and decode the quantum output into a classical form suitable for the target task, such as classification or regression. According to the principles of quantum mechanics, the measurement of a quantum system causes the system to collapse into one of the possible eigenstates of the measured observable. The probability of obtaining a particular eigenvalue is determined by the squared modulus of the corresponding eigenstate's amplitude in the quantum state before measurement.

In QNNs, the observable $O$ is chosen such that its eigenvalues correspond to the desired classical outputs. For example, in a binary classification task, the observable can be defined with eigenvalues $\{0, 1\}$ representing the two classes. The expectation value $\langle O \rangle$ obtained from the QNN circuit, as given in Eq. \ref{eq:eq7}, represents the weighted average of the observable's eigenvalues, where the weights are the probabilities of obtaining each eigenvalue upon measurement.

However, due to the inherent probabilistic nature of quantum mechanics, a single measurement of the quantum state may not provide an accurate estimate of the expectation value. To overcome this limitation, multiple measurements, or ``shots," are performed on identically prepared quantum states.

Let $N$ be the number of shots, and $n_k$ be the number of times the eigenvalue $o_k$ is obtained after $N$ measurements. The estimated probability $\hat{P}(o_k)$ of obtaining the eigenvalue $o_k$ is given by:

\begin{equation}
\label{eq:qc_decoding}
\hat{P}(o_k) = \frac{n_k}{N}.
\end{equation}

The estimated expectation value $\langle \hat{O} \rangle$ can then be computed as:

\begin{equation}
\label{eq:qc_expectation}
\langle \hat{O} \rangle = \sum_k o_k \hat{P}(o_k) = \frac{1}{N} \sum_{i=1}^{N} o_i,
\end{equation}
where $o_i$ is the eigenvalue obtained in the $i$-th measurement.

By performing a large number of shots $N$, the estimated expectation value $\langle \hat{O} \rangle$ converges to the true expectation value $\langle O \rangle$ according to the law of large numbers. The estimated probabilities $\hat{P}(o_k)$ can then be interpreted as the QNN's output for the corresponding classes or regression values, enabling the decoding of the quantum output into a classical form suitable for the target task.

\section{Materials and Methods}
\label{sec:method}

\subsection{Materials}
\subsubsection{Datasets}

\begin{figure}[h]
    \centering
    {\includegraphics[width=\linewidth]{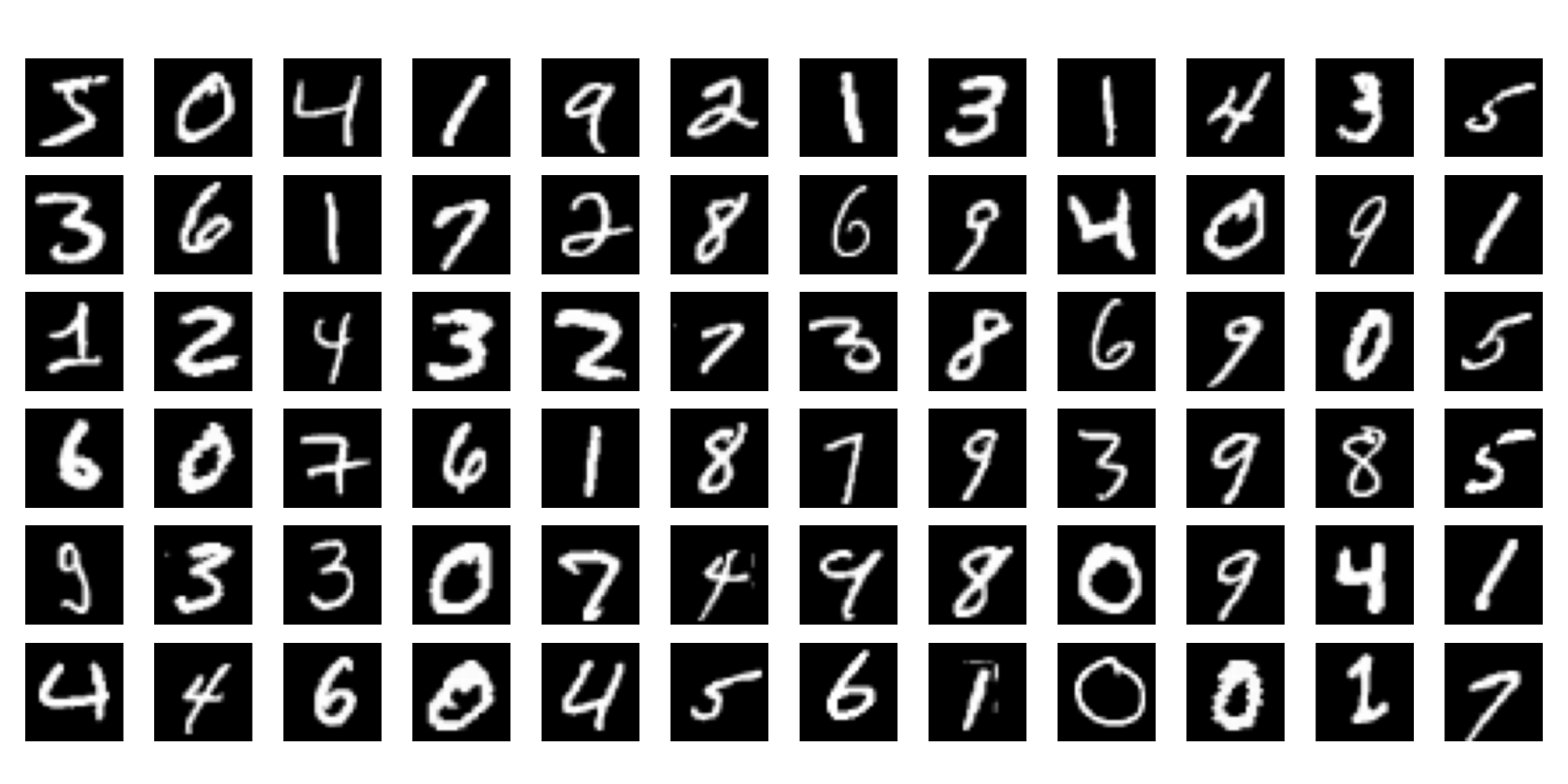}}
    \caption{The MNIST dataset of handwritten digits having 10 classes}
    \label{fig:mnist}
\end{figure}

\begin{figure}[h]
    \centering
    {\includegraphics[width=\linewidth]{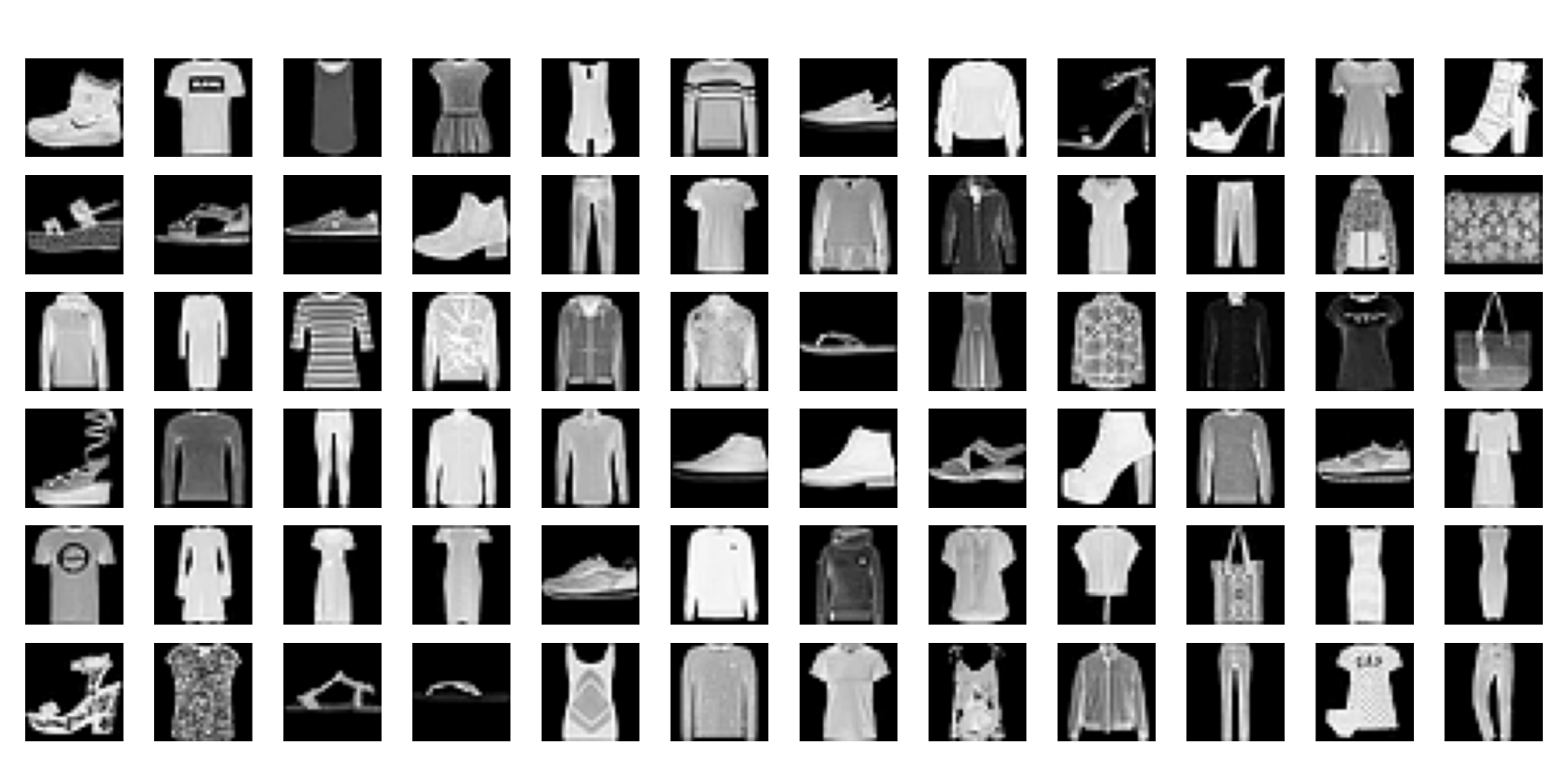}}
    \caption{The FashionMNIST dataset of clothing items with 10 classes}
    \label{fig:fashionmnist}
\end{figure}

\begin{figure}[h]
    \centering
    {\includegraphics[width=\linewidth]{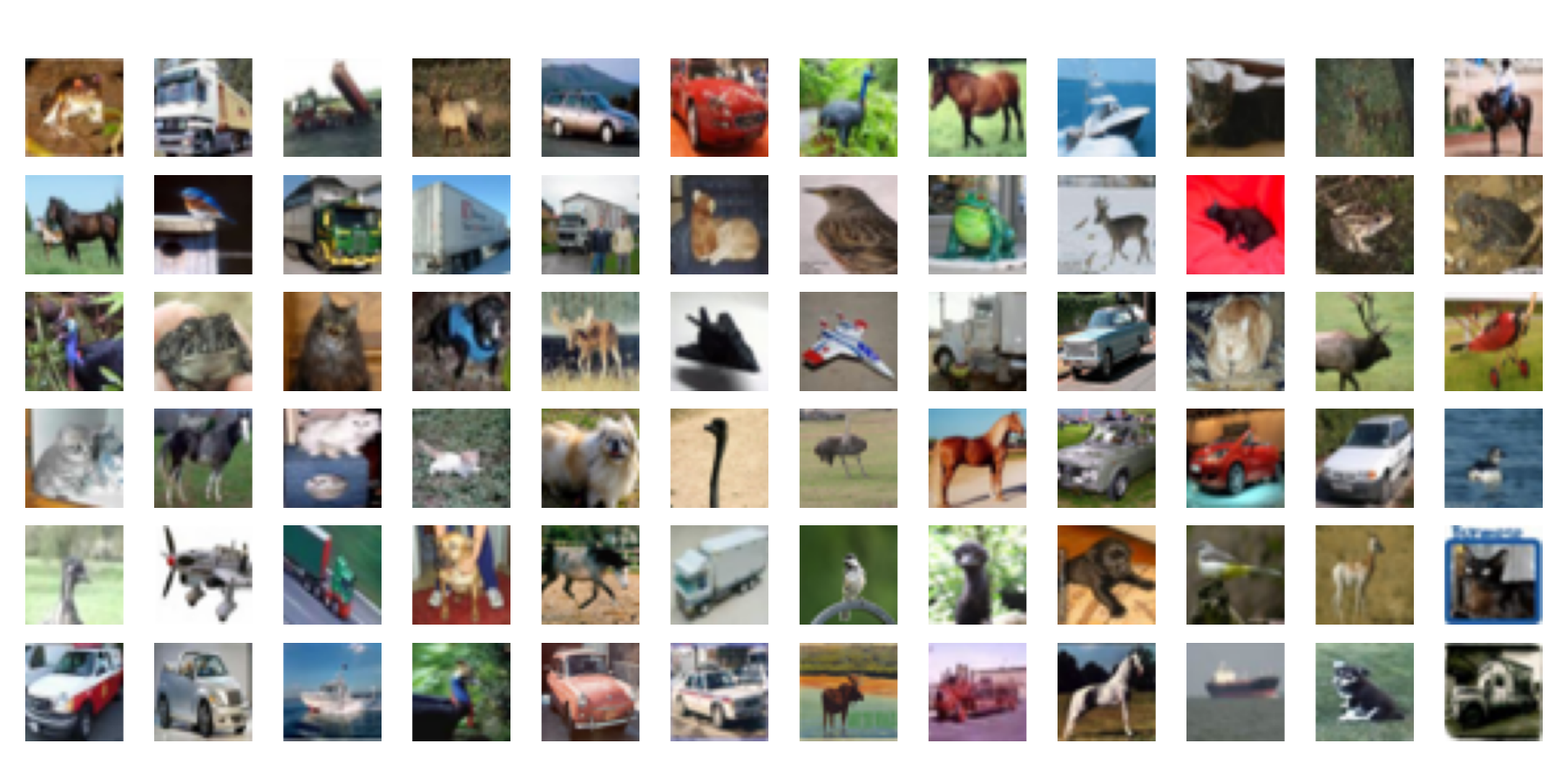}}
    \caption{The CIFAR10 dataset with RGB image classes for image classification}
    \label{fig:cifar10}
\end{figure}

In our experiments, we have employed three well-known datasets: MNIST, FashionMNIST, and CIFAR10. The datasets are shown in Fig. \ref{fig:mnist}, Fig. \ref{fig:fashionmnist}, and Fig. \ref{fig:cifar10}. These datasets are used to demonstrate the generalizability and broader application of the methodology in this study.

{\textbf{MNIST. }}The MNIST dataset \cite{deng2012mnist} is a large database of handwritten digits that is commonly used for training various image processing systems. It contains 60,000 training images and 10,000 testing images, each of which is a grayscale image of size 28$\times$28 pixels. The dataset is designed to allow researchers to benchmark their algorithms in classifying and recognizing handwritten numerical digits from 0 to 9. Due to its simplicity and the fact that it has been extensively studied, it serves as a standard for evaluating the performance of various machine learning techniques.


\textbf{FashionMNIST.} The FashionMNIST dataset \cite{xiao2017fashion} is a more recent dataset comprising 60,000 training samples and 10,000 test samples. Each sample is a 28$\times$28 grayscale image, associated with a label from 10 classes. The classes represent different types of clothing and fashion items, such as T-shirts/tops, trousers, pullovers, dresses, coats, sandals, shirts, sneakers, bags, and ankle boots. This dataset is intended to serve as a direct drop-in replacement for the original MNIST dataset for benchmarking machine learning algorithms, with the added complexity of distinguishing between different types of clothing.

\textbf{CIFAR10. }The CIFAR10 dataset \cite{krizhevsky2009learning} consists of 60,000 32$\times$32 color images in 10 different classes. The classes include airplanes, cars, birds, cats, deer, dogs, frogs, horses, ships, and trucks. There are 6,000 images per class, divided into 50,000 training images and 10,000 test images. The dataset is another benchmark in the field of machine learning and computer vision, providing a more challenging dataset compared to MNIST and FashionMNIST due to its color content and more complex image patterns.

\subsubsection{Classical Convolutional Neural Networks}
Convolutional Neural Networks (CNNs) have been paramount in the field of image recognition and classification. For the MNIST and FashionMNIST datasets, two architectures stand out due to their historical significance and performance: LeNet \cite{lecun1998gradient} and AlexNet \cite{krizhevsky2014one}. In our experiments, we have used these two models as classical teacher models for the grayscale images. Both LeNet and AlexNet were originally intended for higher dimensional images. To accommodate the 28$\times$28 pixel grayscale images of MNIST and FashionMNIST, we have modified the input layers of these models to accept single-channel 28$\times$28 pixel input instead of their original input dimensions.

For the CIFAR10 dataset, which consists of 32$\times$32 pixel color images, we have utilized pre-trained models of ResNet50 \cite{he2016deep}, DenseNet121 \cite{huang2017densely}, and EfficientNetB0 \cite{tan2019efficientnet} without any architectural modifications. These models are well-suited for handling the higher complexity and information present in the CIFAR10 images.

\begin{figure*}[t]
    \centering
    {\includegraphics[width=1\linewidth, height = 11em]{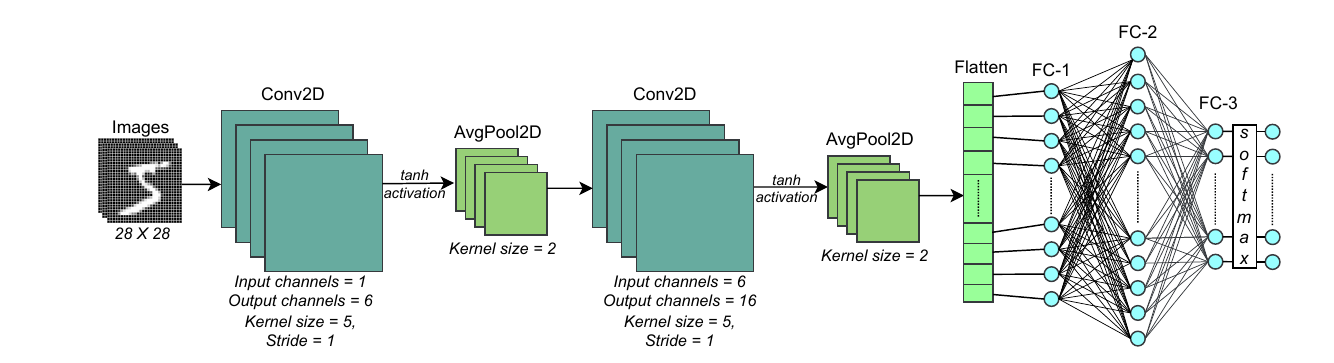}}
    \caption{The modified LeNet architecture used in the implementation of the experiments. Two convolution layers, each followed by tanh activation and average pool layers were used. Finally, three fully connected layers were used to classify the images and get output probabilities from the model}
    \label{fig:lenet}

    \vspace{1em}
    {\includegraphics[width=1\linewidth, height = 11em]{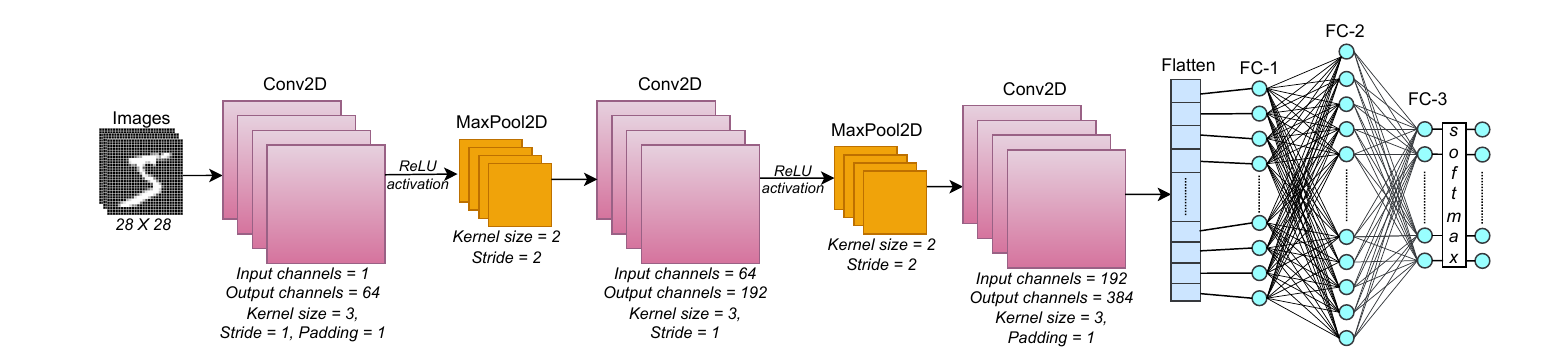}}
    \caption{The modified AlexNet architecture used in the implementation of the experiments. Three convolution layers, first two followed by ReLU activation and max pool layers were used. Finally, three fully connected layers were used to classify the images and get output probabilities from the model}
    \label{fig:alexnet}
\end{figure*}

\textbf{Modified LeNet Architecture.} LeNet is one of the earliest convolutional neural networks that significantly impacted the field of deep learning. It is a much simpler network compared to AlexNet, consisting of two convolutional layers followed by two fully connected layers. LeNet is particularly well-suited for handwritten digit recognition tasks like MNIST due to its architecture being designed for this purpose. In our experiments, we have used the original LeNet architecture by changing the input layer to accept images of dimension 28$\times$28 instead of its original input image dimension of 32$\times$32, as shown in Fig. \ref{fig:lenet}. The intermediate layers were constructed according to the original architecture with no change.

\textbf{Modified AlexNet Architecture.} AlexNet is a deep CNN that marked a breakthrough in the ImageNet competition. It consists of five convolutional layers followed by three fully connected layers. AlexNet takes an RGB image of size 256$\times$256 as input and uses 11$\times$11 filters with a stride of 4 in the first convolutional layer. Since MNIST and FashionMNIST images are grayscale images of size 28$\times$28, the architecture of AlexNet was modified in the intermediate layers. Fig. \ref{fig:alexnet} shows the followed architecture of AlexNet that was used instead: The modified architecture features an initial convolutional layer with 64 filters of size 3$\times$3, stride 1, and padding 1, followed by ReLU activation and a max pooling layer with a 2$\times$2 kernel and stride 2. The second layer has 192 filters of size 3$\times$3 with padding 1, followed by ReLU activation and another 2$\times$2 max pooling. The third convolutional layer consists of 384 filters with a kernel size of 3$\times$3 and padding of 1, followed by ReLU activation. This is followed by a series of fully connected layers: the first with 4096 neurons, the second also with 4096 neurons, and the final layer corresponding to the number of classes. Dropout is applied before the first and second fully connected layers to prevent overfitting. The model is adapted for single-channel grayscale images and the smaller spatial dimensions of the MNIST and FashionMNIST datasets. The use of ReLU activation functions, dropout, and overlapping pooling makes AlexNet robust against overfitting and capable of learning complex patterns in image data. In the context of MNIST and FashionMNIST, AlexNet's architecture, although more complex than necessary for such simple datasets, provides an excellent baseline for performance due to its depth and capacity to learn intricate features.



\begin{figure}[ht!]
    \centering
    {\includegraphics[width=0.7\linewidth]{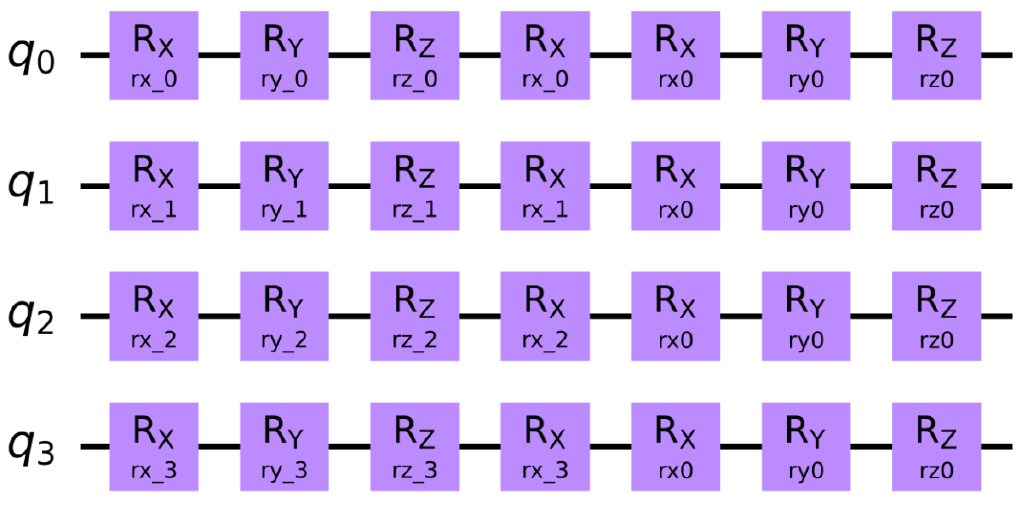}}
    \caption{The 4-qubit PQC with Rx, Ry and Rz gates parameterized by $\theta$ used as student model for image classification}
    \label{fig:qc1}
    \centering
    {\includegraphics[width=0.7\linewidth]{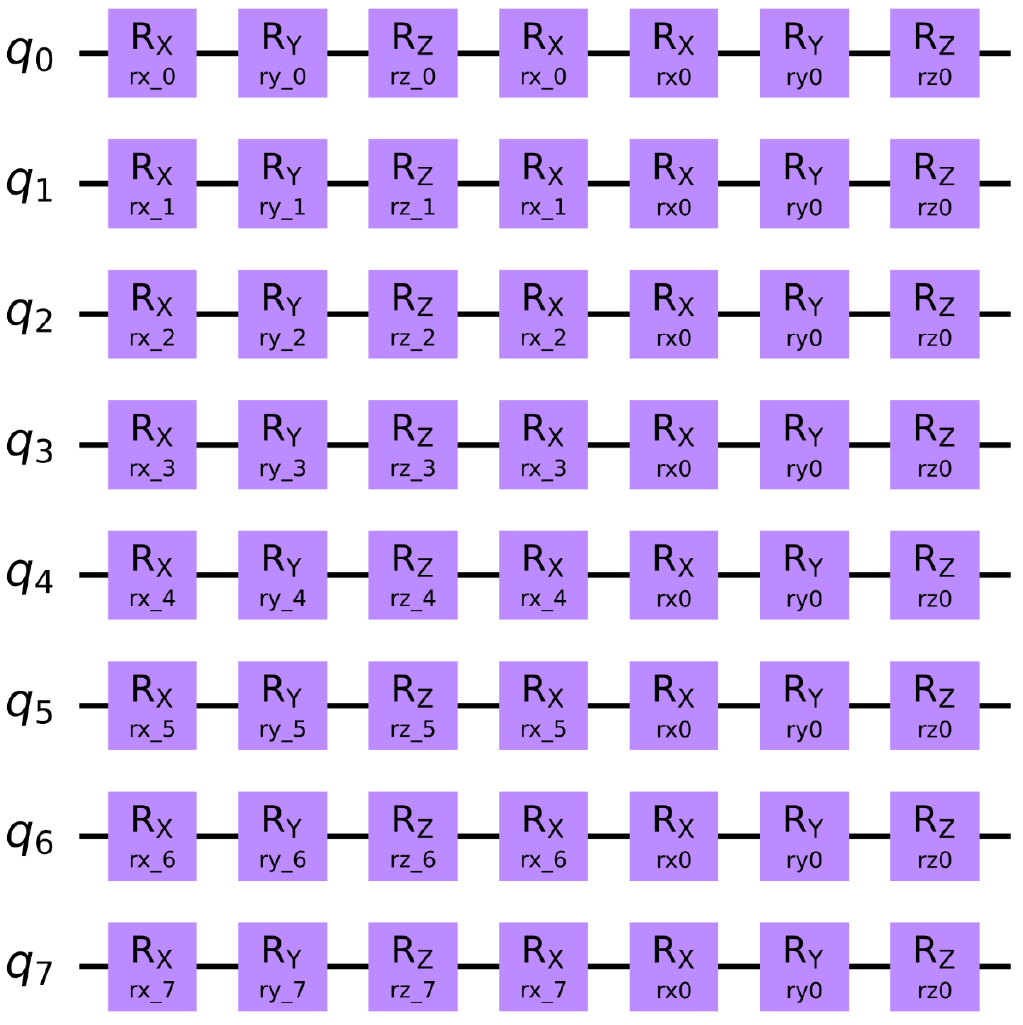}}
    \caption{The 8-qubit PQC with Rx, Ry and Rz gates parameterized by $\theta$ used as student model for image classification}
    \label{fig:qc2}
\end{figure}

\subsubsection{Parameterized Quantum Circuits}

In our study, we employed two Parameterized Quantum Circuits (PQCs): one with 4 qubits and another with 8 qubits. PQCs are central to QML and variational algorithms, where the parameters (often denoted as \( \theta \)) of quantum gates are tuned to minimize a cost function. The PQCs that are constructed and used in our experiments are shown in Fig. \ref{fig:qc1} and Fig. \ref{fig:qc2}. The PQCs are constructed using rotation gates Rx, Ry and Rz, which are parameterized by angles \( \theta \). These angles are randomly initialized and optimized during the learning process. The 4-qubit circuit provides a compact yet expressive model that can capture the complexity needed for smaller-scale quantum tasks. With four additional qubits, the 8-qubit circuit can represent a larger state space, allowing it to capture more complex patterns and correlations in the data.


\subsubsection{Performance Metrics}

To evaluate the performance of our quantum and classical models, we rely on prediction accuracy as the primary metric. Accuracy is a suitable measure in the context of balanced datasets, such as the ones we have crafted through our filtration process. The accuracy of a model is calculated as the ratio of correctly predicted instances to the total number of instances, as shown in Eq. \ref{eq:eq14}:

\begin{equation}
\label{eq:eq14}
\text{Accuracy} = \frac{\text{Number of correct predictions}}{\text{Total number of predictions}}.
\end{equation}

Since our filtered datasets are balanced with an equal number of samples from each class, accuracy serves as a reliable metric. In datasets where classes are imbalanced, metrics such as precision, recall, or the F1 score might be more appropriate. However, for balanced datasets, accuracy straightforwardly reflects the model's capability to classify new data points correctly. Accuracy is not only a clear and interpretable metric but also practical from a computational standpoint. It allows us to clearly demonstrate the effectiveness of our models without incurring additional computational costs or complexity. Accuracy provides a direct measure of a model's classification performance and is a widely used metric in machine learning research for balanced datasets, making it an appropriate choice for this study.

\subsubsection{Implementation Details}
In our experimental setup, we employed PyTorch \cite{paszke2019pytorch} as the framework for implementing neural network models, with TorchQuantum library \cite{hanruiwang2022quantumnas} used for conducting quantum simulations. Hyperparameter optimization was used, enabling the selection of optimal values for the optimizer, learning rate, loss function, temperature for KL-Divergence, and alpha ($\alpha$) (Eq. \ref{eq:eq3}). The Adam optimizer \cite{kingma2014adam} was chosen for its effectiveness in handling noisy problems and sparse gradients. We set the learning rate to 0.001, and run all experiments for 1000 epochs with early stopping regularizer with a patience = 10 epochs to prevent overfitting while allowing sufficient model training. The loss function employed was Cross-Entropy Loss \cite{zhang2018generalized} for classification, complemented by KL-Divergence for the knowledge distillation process. The temperature \( \tau \) and loss weight $\alpha$ was found to be \( \tau \) = 2 and $\alpha$ = 0.4. For the quantum experiments, we perform the default 1024 shots set by TorchQuantum for each measurement to approximate the expectation values with high accuracy and interpret them as probability distributions. \textbf{We perform each experiment five times and report the mean of the results with the standard deviation to make them more statistically meaningful.}

\subsection{Proposed Methodology}
\subsubsection{Problem Formulation}

We consider a classical neural network as the teacher model and a quantum neural network (QNN) as the student model. The teacher model is denoted by \( f_t: \mathcal{X} \rightarrow \mathcal{Y} \), where \( \mathcal{X} \) is the input space and \( \mathcal{Y} \) is the output space. Similarly, the student model is denoted by \( f_s: \mathcal{X} \rightarrow \mathcal{Y} \).

Given a dataset \( \mathcal{D} = \{(\mathbf{x}_i, y_i)\}_{i=1}^N \) where \( \mathbf{x}_i \in \mathcal{X} \) and \( y_i \in \mathcal{Y} \), the task is to train the student QNN to approximate the mapping of the teacher model. The performance of the QNN can be evaluated by a loss function \( \mathcal{L}: \mathcal{Y} \times \mathcal{Y} \rightarrow \mathbb{R} \), which measures the discrepancy between the predictions of the QNN and the ground truth labels. The objective function for the QNN is given in Eq. \ref{eq:eq1} as:
\begin{equation}
\label{eq:eq1}
\mathcal{O}(f_s) = \frac{1}{N} \sum_{i=1}^N \mathcal{L}(f_s(\mathbf{x}_i), y_i).
\end{equation}

The loss of the model outputs, $\mathcal{L}$, is calculated using cross-entropy loss \cite{zhang2018generalized} according to Eq. \ref{eq:15}:
\begin{equation}
\label{eq:15}
\mathcal{L}_{\text{cross-entropy}}(y, \hat{y}) = -\sum_{i=1}^{C} y_i \log(\hat{y}_i)
\end{equation}
where \( C \) is the total number of classes, \( y_i \) is the ground truth label indicating the probability of the predicted class, and \( \hat{y}_i \) is the predicted probability that the observation is of class \( i \).

\begin{figure*}[t]
    \centering
    {\includegraphics[width=\linewidth]{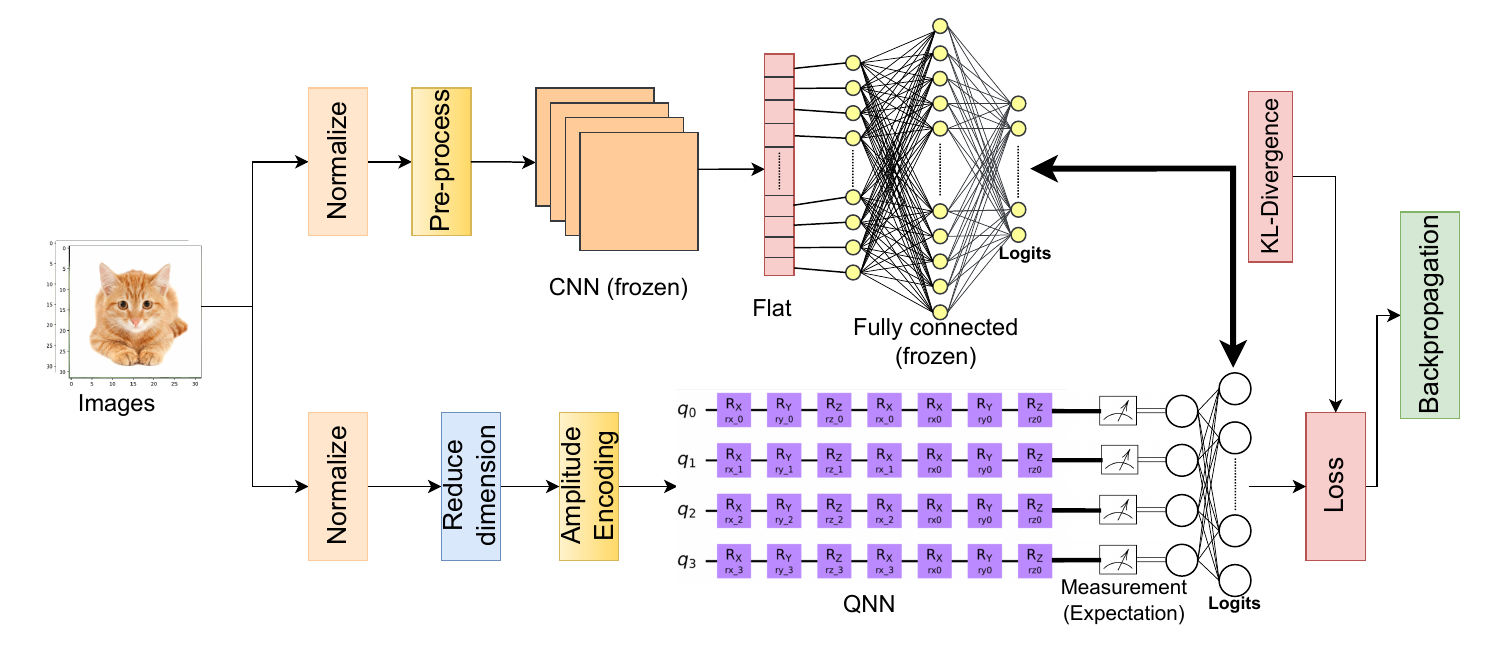}}
    \vspace*{-1.5em}
    \caption{The architecture diagram of the research methodology followed in this research. Two networks, a classical frozen network with CNN and fully connected layers, and a trainable quantum network with data encoder and QNN is used. The input images are normalized and pre-processed respectively for the classical CNN model and the quantum QNN model. Following that, the data is passed through the respective models. Outputs from the CNN are passed through a fully connected layer to get output logits. Measurements on the qubits of the QNN gives expectation values (or quantum logits). The logits from the classical network and the QNN are used to calculate the KL-divergence, which is added to the loss of the quantum network to aid in backpropagation}
    \label{fig:workflow}
\end{figure*}

The methodology followed in this study is visualized in Fig. \ref{fig:workflow}. The figure shows that after following data pre-processing steps for classical and quantum models respectively, the classical data is passed through a classical network containing teacher CNN model(s), whereas the quantum data is passed through student QNN model(s). After that, using the outputs of the models, the KL-Divergence is calculated and added to the loss of the student model to calculate a total loss which is used to improve performance of the models. Adding the KL-divergence acts as a supervisory signal while updating the parameters of the quantum model. In this teacher-student setup, the student QNN thus tries to mimic the outputs of the teacher CNN and knowledge transfer is accomplished from classical to quantum models solely based on the mimicry. We discuss the used methods in each step of our experiments in the following portion.

\subsubsection{Data Normalization}
Data normalization is a crucial preprocessing step in many machine learning tasks, including classical and quantum neural networks. Normalization ensures that the input features are on a similar scale, which can improve the convergence rate and generalization performance of the models.

In this study, we employ a widely used normalization technique known as standardization or z-score normalization. This method rescales the features to have a mean of 0 and a standard deviation of 1. For a feature vector $\mathbf{x} = (x_1, x_2, \ldots, x_n)$, the standardized feature vector $\mathbf{z} = (z_1, z_2, \ldots, z_n)$ is obtained as:

\begin{equation}
\label{eq:z_score}
z_i = \frac{x_i - \mu_i}{\sigma_i},
\end{equation}
where $\mu_i$ and $\sigma_i$ are the mean and standard deviation of the $i$-th feature, respectively, computed across the entire dataset.

The mean $\mu_i$ and standard deviation $\sigma_i$ are calculated as:

\begin{equation}
\label{eq:mean}
\mu_i = \frac{1}{N} \sum_{j=1}^{N} x_{ij},
\end{equation}

\begin{equation}
\label{eq:std_dev}
\sigma_i = \sqrt{\frac{1}{N} \sum_{j=1}^{N} (x_{ij} - \mu_i)^2},
\end{equation}
where $N$ is the total number of samples in the dataset, and $x_{ij}$ is the $i$-th feature of the $j$-th sample.

Standardization ensures that the features are centered around 0 and have a uniform scale, preventing any single feature from dominating the others due to its original scale. This normalization technique is particularly beneficial when dealing with heterogeneous features or when the input features have different units or ranges.

For image data, which is the focus of this study, normalization is typically performed by rescaling the pixel values to a specific range, such as [0, 1] or [-1, 1]. This can be achieved by dividing the pixel values by the maximum possible value (e.g., 255 for 8-bit images).

The mean and standard deviation used for normalization are computed on the training set and applied consistently to the validation and test sets during inference. This ensures that the normalization parameters are not influenced by the unseen data, maintaining the integrity of the evaluation process.

\subsubsection{Classical Data Pre-processing and Quantum Data Encoding}
\label{sec:encode_method}
For neural networks, the data pre-processing steps involve converting the raw data into a suitable format for efficient processing. In this study, we work with image datasets, and the following steps are performed:

\begin{enumerate}
    \item \textbf{Conversion to Tensors}: The raw image data is converted into PyTorch tensors, which are multi-dimensional arrays that facilitate efficient computations on GPUs or other hardware accelerators.
    \item \textbf{Dataset Creation}: The tensor data is organized into a PyTorch Dataset, which provides a convenient way to access and iterate over the data samples during training and evaluation.
    \item \textbf{DataLoader Creation}: PyTorch DataLoaders are created from the Datasets, enabling efficient batching, shuffling, and parallel loading of data during the training process.
\end{enumerate}

These pre-processing steps ensure that the data is properly formatted and optimized for efficient processing by the classical neural network models. After the pre-processing steps, for QNNs, we use angle amplitude encoding to encode the classical data into quantum states. Amplitude encoding is a suitable choice for QNNs as it directly maps the classical data into the amplitudes of the quantum states. In amplitude encoding, it is possible to encode $2^Q$ features into $Q$ qubits. This is particularly useful in the NISQ era since the number of available qubits is very low in the quantum systems. It is advantageous when dealing with high-dimensional data or complex feature spaces, as the exponential state space of quantum systems can provide a compact and expressive representation. Moreover, amplitude encoding naturally accommodates both continuous and discrete data, eliminating the need for discretization or binarization steps that may lead to information loss. By directly encoding data into the amplitudes of quantum states, QNNs can potentially exploit quantum parallelism and quantum interference effects to perform computations more efficiently than their classical counterparts.

\subsubsection{Dimensionality Reduction for Quantum Circuits}
\label{sec:dimen_reduc}
Quantum circuits operate on qubits, which have a limited number of dimensions. Consequently, when working with high-dimensional classical data, such as images, it is necessary to reduce the dimensionality of the data before encoding it into the quantum state. In this study, we employ a dimensionality reduction technique that utilizes a fully connected neural network with two layers.

Let $Q$ denote the number of qubits in our quantum circuits, which is 4 or 8 in our experiments. The dimensionality of the classical data needs to be reduced to $2^Q$ to be encoded into $Q$ qubits. The dimensionality reduction process is as follows:

1. \textbf{Flattening}: The input image is flattened into a one-dimensional vector, effectively converting the two-dimensional image into a high-dimensional vector representation.

2. \textbf{Fully Connected Layer 1}: The flattened vector is fed into the first fully connected layer, which acts as a non-linear transformation, mapping the high-dimensional input to an intermediate representation.

3. \textbf{Fully Connected Layer 2}: The second fully connected layer takes the output of the first layer and reduces the dimensionality to $2^Q$, producing a $2^Q$-dimensional vector that can be directly encoded into the quantum state using angle encoding.

The use of a fully connected neural network for dimensionality reduction offers several advantages over other techniques, such as center cropping, Principal Component Analysis (PCA), or pooling operations like maximum pooling or average pooling. The fully connected layers can learn an optimized transformation that captures the most relevant features from the input data, tailored to the specific task and dataset. In addition, the non-linear activation functions in the fully connected layers allow the network to model complex, non-linear relationships between the input and output dimensions, which may be beneficial for representing the intricate patterns present in the data. Moreover, by incorporating the dimensionality reduction layers within the quantum neural network architecture, the parameters of these layers can be optimized jointly with the rest of the quantum circuit during the training process, potentially leading to better overall performance.

\subsubsection{Integration of Knowledge Distillation}
Knowledge Distillation aims to transfer knowledge from the teacher model to the student model by minimizing a distillation loss \( \mathcal{L}_{KD} \). The process involves a soft target \( \tau \) which controls the smoothness of the teacher’s output probability distribution. The distillation loss can be defined using the Kullback-Leibler (KL) divergence \cite{kullback1951information} in Eq. \ref{eq:eq2} as:
\begin{equation}
\label{eq:eq2}
\mathcal{L}_{KD}(f_s, f_t) = \frac{1}{N} \sum_{i=1}^N KL\Big(\sigma\big(\frac{f_t(\mathbf{x}_i)}{\tau}\big) \ || \ \sigma\big(\frac{f_s(\mathbf{x}_i)}{\tau}\big)\Big),
\end{equation}
where \( \sigma \) denotes the softmax function and KL denotes the Kullback-Leibler divergence. The softmax function is given according to Eq. \ref{eq:softmax} by:
\begin{equation}
\label{eq:softmax}
\sigma(\mathbf{z})_i = \frac{e^{z_i}}{\sum_{j=1}^{K} e^{z_j}}
\end{equation}
where \( \sigma(\mathbf{z})_i \) is the output of the softmax function for the \( i \)-th class, \( z_i \) is the \( i \)-th element of the input vector \( \mathbf{z} \), and \( K \) is the total number of classes. The exponential function \( e^{z_i} \) ensures that all outputs are non-negative and the denominator normalizes the values to ensure they sum up to 1, thus forming a valid probability distribution.

And, the KL-divergence is calculated as per Eq. \ref{eq:kl_divergence}:
\begin{equation}
\label{eq:kl_divergence}
KL(P || Q) = \sum_{i=1}^{K} P(i) \log\left(\frac{P(i)}{Q(i)}\right)
\end{equation}
where \( P \) and \( Q \) are discrete probability distributions. \( P(i) \) represents the probability of the \( i \)-th class according to the true label distribution, typically obtained from the softmax function of the teacher model \( \sigma(f_t(\mathbf{x}_i)) \), and \( Q(i) \) is the probability of the \( i \)-th class according to the predicted label distribution from the student model \( \sigma(f_s(\mathbf{x}_i)) \). \( K \) is the total number of classes. The KL divergence measures the difference between two probability distributions over the same variable. The total loss for the student QNN, incorporating both the original objective and the distillation loss, is given in Eq. \ref{eq:eq3} by:
\begin{equation}
\label{eq:eq3}
\mathcal{O}_{total}(f_s) = \alpha\mathcal{L}_{KD}(f_s, f_t) + (1 - \alpha)\mathcal{O}(f_s),
\end{equation}
where \( \alpha \) is a hyperparameter that balances the importance of the original loss and the distillation loss. In order to balance the supervision from the distillation loss, the overall loss is balanced as $0 \leq \alpha \leq 1$.

The student QNN is trained by optimizing \( \mathcal{O}_{total} \) through a quantum optimization algorithm. The goal of the optimization is to minimize the loss, which can be formulated in Eq. \ref{eq:eq4} as:
\begin{equation}
\label{eq:eq4}
f_s^* = \arg\min_{f_s} \mathcal{O}_{total}(f_s).
\end{equation}

\section{Results and Analysis}
\subsection{Preliminary Results}
\begin{figure}[t!]
    \centering
    {\includegraphics[width=0.6\linewidth]{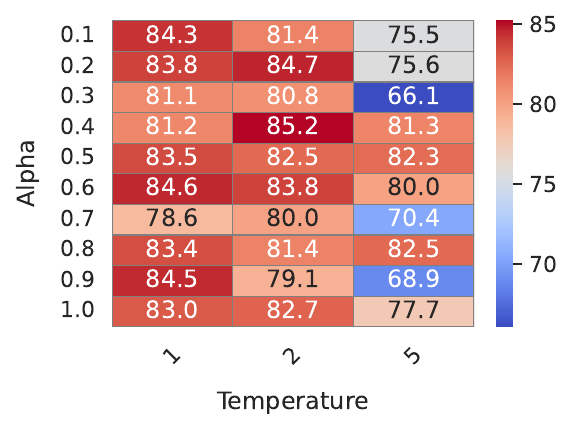}}
    \caption{Performance of the 4-qubit PQC student model with different values of hyperparameters}
    \label{fig:hyper}
\end{figure}
Initially, we set up experiments to find the best hyperparameters. In the case of knowledge distillation, hyperparameters \( \tau \) and $\alpha$ play the most important role in the model performance. Thus, we create the hyperparameter space \( \tau \) = [1, 2, 5] and $\alpha$ = [0.1, 0.2, 0.3, 0.4, 0.5, 0.6, 0.7, 0.8, 0.9, 1.0], and perform experiments by dividing the training sets into training and validation sets. This helped us to select the optimal values of temperature (\( \tau \)) and alpha (\( \alpha \)), which are crucial for the knowledge distillation process. The weight of the distillation is represented by \( \alpha \), while \( 1-\alpha \) represents the weight of the main training loss. Fig. \ref{fig:hyper} shows a heatmap of the performance of the student model (4-qubit PQC) with varying \( \alpha \) and \( \tau \) values, using AlexNet as the teacher model.

From the heatmap, we observe that the student model benefits from smaller values of \( T \) and \( \alpha \) in general. Specifically, lower temperatures tend to sharpen the probability distribution, which helps the student model learn more confidently from the teacher's predictions. This signifies that the student model benefits from a more focused and less noisy learning signal. Similarly, a smaller \( \alpha \) indicates a lower reliance on the distilled knowledge from the teacher, which seems to improve the student's performance. This implies that the student model can effectively leverage the teacher's knowledge to achieve higher accuracy with lower weightage on supervisory signals through the distillation process. Concurrently, it implies that the student model possesses a certain degree of inherent capability or intuition that allows it to perform well even without heavily relying on the distilled knowledge from the teacher. This inherent capability might be attributed to the student model's structure, its initial conditions, or its ability to generalize from limited information. The observed trend suggests that while knowledge distillation helps in improving the student model's performance, overly depending on the teacher's output could potentially constrain the student model's ability to explore and learn from the data itself.

\subsection{Main Results}
\begin{table*}[!htbp]
\centering
\caption{Performance of quantum student models and classical teacher models on the MNIST, FashionMNIST, and CIFAR10 datasets, with and without knowledge distillation. T represents teacher and S represents the student model in distillation experiments.}
\label{tab:combined}
\begin{tabular}{l l l c}
\toprule
\textbf{Dataset} & \textbf{Category} & \textbf{Architecture} & \textbf{Accuracy (\%)} \\
\midrule
\multirow{8}{*}{MNIST} 
  & \multirow{2}{*}{Student} 
      & 4-qubit            & $77.69 \pm 0.25$ \\
  &                      & 8-qubit            & $90.40 \pm 0.15$ \\
\cmidrule(lr){2-4}
  & \multirow{2}{*}{Teacher} 
      & AlexNet            & $99.44 \pm 0.03$ \\
  &                      & LeNet              & $98.99 \pm 0.05$ \\
\cmidrule(lr){2-4}
  & \multirow{4}{*}{Distillation} 
      & T: AlexNet, S: 4-qubit  & \cellcolor{green!25}$85.18 \pm 0.20$ (+7.49) \\
  &                      & T: LeNet, S: 4-qubit    & \cellcolor{green!25}$84.07 \pm 0.18$ (+6.38) \\
  &                      & T: AlexNet, S: 8-qubit  & \cellcolor{green!25}$91.27 \pm 0.12$ (+0.87) \\
  &                      & T: LeNet, S: 8-qubit    & \cellcolor{green!25}$90.63 \pm 0.10$ (+0.23) \\
\midrule
\multirow{8}{*}{FashionMNIST} 
  & \multirow{2}{*}{Student} 
      & 4-qubit            & $77.36 \pm 0.30$ \\
  &                      & 8-qubit            & $83.44 \pm 0.20$ \\
\cmidrule(lr){2-4}
  & \multirow{2}{*}{Teacher} 
      & AlexNet            & $93.68 \pm 0.08$ \\
  &                      & LeNet              & $90.00 \pm 0.12$ \\
\cmidrule(lr){2-4}
  & \multirow{4}{*}{Distillation} 
      & T: AlexNet, S: 4-qubit  & \cellcolor{green!25}$80.61 \pm 0.25$ (+3.25) \\
  &                      & T: LeNet, S: 4-qubit    & \cellcolor{green!25}$80.49 \pm 0.22$ (+3.13) \\
  &                      & T: AlexNet, S: 8-qubit  & \cellcolor{green!25}$85.23 \pm 0.18$ (+1.79) \\
  &                      & T: LeNet, S: 8-qubit    & \cellcolor{green!25}$84.71 \pm 0.15$ (+1.27) \\
\midrule
\multirow{11}{*}{CIFAR10} 
  & \multirow{2}{*}{Student} 
      & 4-qubit            & $25.44 \pm 0.40$ \\
  &                      & 8-qubit            & $31.62 \pm 0.35$ \\
\cmidrule(lr){2-4}
  & \multirow{3}{*}{Teacher} 
      & ResNet50           & $83.38 \pm 0.15$ \\
  &                      & DenseNet121        & $88.24 \pm 0.12$ \\
  &                      & EfficientNetB0     & $87.38 \pm 0.10$ \\
\cmidrule(lr){2-4}
  & \multirow{6}{*}{Distillation} 
      & T: ResNet50, S: 4-qubit       & \cellcolor{green!25}$29.13 \pm 0.35$ (+3.69) \\
  &                      & T: DenseNet121, S: 4-qubit    & \cellcolor{green!25}$29.05 \pm 0.32$ (+3.61) \\
  &                      & T: EfficientNetB0, S: 4-qubit & \cellcolor{green!25}$27.32 \pm 0.28$ (+1.88) \\
  &                      & T: ResNet50, S: 8-qubit       & \cellcolor{green!25}$34.04 \pm 0.30$ (+2.42) \\
  &                      & T: DenseNet121, S: 8-qubit    & \cellcolor{green!25}$34.22 \pm 0.28$ (+2.60) \\
  &                      & T: EfficientNetB0, S: 8-qubit & \cellcolor{green!25}$32.87 \pm 0.25$ (+1.25) \\
\bottomrule
\end{tabular}
\end{table*}
We present the experimental results on the three datasets in Table \ref{tab:combined}. From the results, we can observe several patterns and trends. \textbf{\textit{(i)}} The knowledge distillation process from classical to quantum models demonstrates substantial gains in accuracy across all distillation experiments. We observe an average accuracy improvement of 3.74\% on the MNIST, 2.36\% on the FashionMNIST, and 2.58\% on the CIFAR10 dataset. This shows that knowledge distillation from classical to quantum neural networks is not only possible, but can be applied across different datasets and models. \textbf{\textit{(ii)}} The performance of QNN models as students shows a marked improvement when the number of qubits is increased from 4 to 8 across all datasets. This suggests that the capacity of QNNs scales significantly with the number of qubits, which is a promising sign for the scalability of quantum models. \textbf{\textit{(iii)}} The distillation benefits are more prominent for smaller quantum models. The 4-qubit student models gain more in terms of accuracy percentage points compared to the 8-qubit models when distilled from the same teacher architecture. This could indicate that smaller QNNs are more receptive to the distilled knowledge, possibly due to their lower initial performance providing a greater scope for improvement. \textbf{\textit{(iv)}} The ResNet model, despite having lower baseline accuracy than DenseNet and EfficientNet, improves the accuracy of the student models consistently more than EfficientNet and also more than DenseNet for the 4-qubit model. This can be attributed to the architectural characteristics of ResNet. ResNet's design incorporates residual blocks with skip connections that facilitate the flow of gradients during training. This allows the student models to benefit from deeper feature hierarchies without the hindrance of vanishing gradients, which is particularly advantageous for smaller models like the 4-qubit QNNs. In contrast, EfficientNet relies on compound scaling and a balance of network width, depth, and resolution for efficient performance. While this makes EfficientNet highly efficient, it may not provide the same level of rich feature extraction as ResNet, which is crucial for knowledge distillation. DenseNet, with its densely connected architecture, encourages feature reuse and reduces the number of parameters. However, the direct connections between all layers may not be as effective in transferring knowledge to a quantum model due to the potential complexity and noise in the feature maps being transferred. Therefore, the geometric properties of ResNet, which simplify and enhance the representational capacity of the data, make it a more suitable teacher for the student models in the context of knowledge distillation. \textbf{\textit{(v)}} The teacher models' performance serves as a high benchmark for the student QNNs. While the student models do not yet match the teachers' accuracy, the gap narrows significantly post-distillation. This narrowing gap is indicative of the potential for QNNs to approach the performance of classical models, given sufficient advancements in quantum computing and model design in the future.

\subsection{Statistical Significance}

To validate the improvements in accuracy obtained through knowledge distillation, we conducted paired t-tests to assess the statistical significance of our results. The p-values for the comparisons between student models with and without knowledge distillation are presented in Table~\ref{tab:pvalues}. 

\begin{table}[!htbp]
\centering
\caption{P-values for paired t-tests on the performance improvements due to knowledge distillation.}
\label{tab:pvalues}
\begin{tabular}{lcc}
\toprule
\textbf{Dataset} & \textbf{Model} & \textbf{P-value} \\
\midrule
MNIST & 4-qubit & $1.11 \times 10^{-39}$ \\
 & 8-qubit & $3.61 \times 10^{-18}$ \\
\midrule
FashionMNIST & 4-qubit & $2.89 \times 10^{-26}$ \\
 & 8-qubit & $2.29 \times 10^{-23}$ \\
\midrule
CIFAR10 & 4-qubit & $6.58 \times 10^{-24}$ \\
 & 8-qubit & $3.51 \times 10^{-20}$ \\
\bottomrule
\end{tabular}
\end{table}

The results demonstrate that the accuracy improvements obtained through knowledge distillation are statistically significant across all datasets and model architectures, with p-values significantly less than 0.05. This confirms the effectiveness of our knowledge distillation approach from classical to quantum neural networks.

\subsection{Ablation Studies}
In this part, we conduct several ablation studies to understand the contribution and effect of the selected methodologies to the overall performance of the student models. To this goal, we conduct ablation studies on the dimensionality reduction techniques and the data encoding techniques applied to the student model, and show a comparative analysis of the performance of the models, as well as show that the best methods were selected. 

\subsubsection{Ablation on Dimensionality Reduction Techniques}

\begin{figure}[!htbp]

\centering
\vspace*{-2em}  
\begin{subfigure}{\textwidth}
  \centering
  \vspace*{-2em}  \includegraphics[width=1\linewidth]{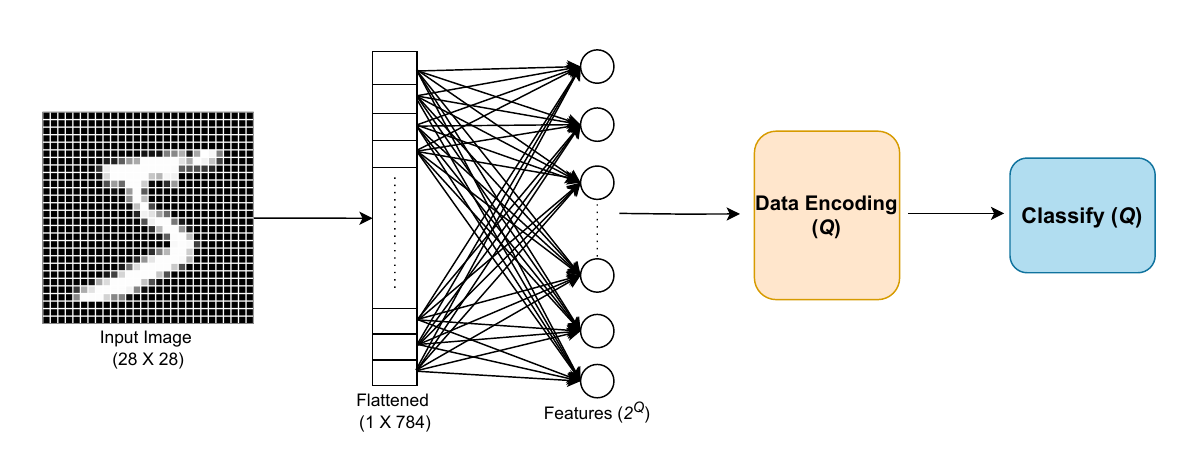}
  \vspace*{-2em}  
  \caption{} 
  \label{fig:sub1}
\end{subfigure}%


\begin{subfigure}{\textwidth}
  \centering
 \includegraphics[width=1\linewidth]{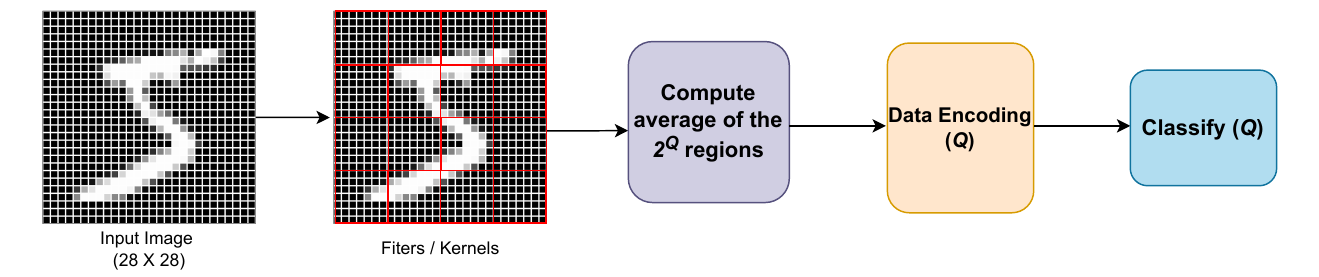}
\vspace*{-2em}
  \caption{}
  \label{fig:sub2}
\end{subfigure}%

\begin{subfigure}{\textwidth}
  \centering
  \includegraphics[width=1\linewidth]{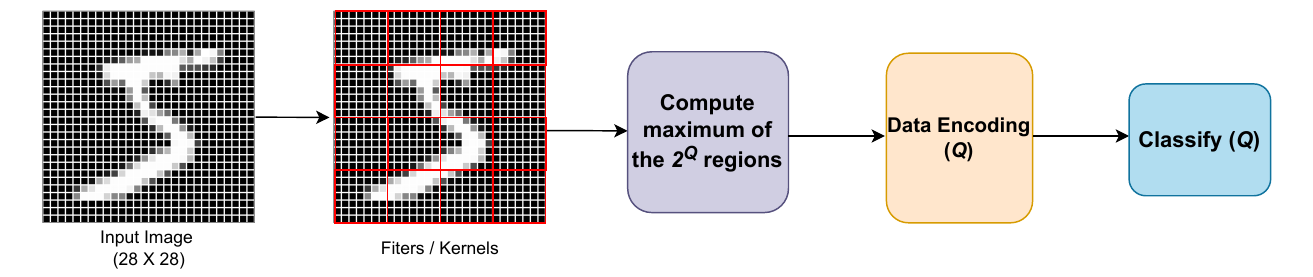}
  \vspace*{-2em}
  \caption{}
  \label{fig:sub3}
\end{subfigure}%

\begin{subfigure}{\textwidth}
  \centering
  \includegraphics[width=1\linewidth]{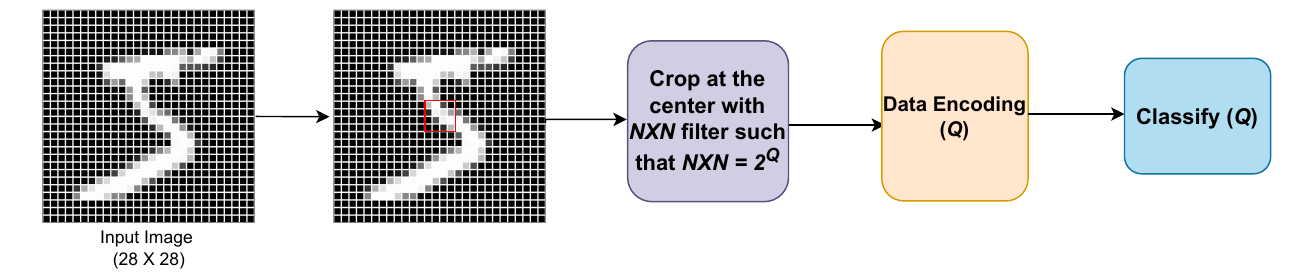}
  \vspace*{-2em}
  \caption{}
  \label{fig:sub4}
\end{subfigure}%

\begin{subfigure}{\textwidth}
  \centering
    \includegraphics[width=1\linewidth]{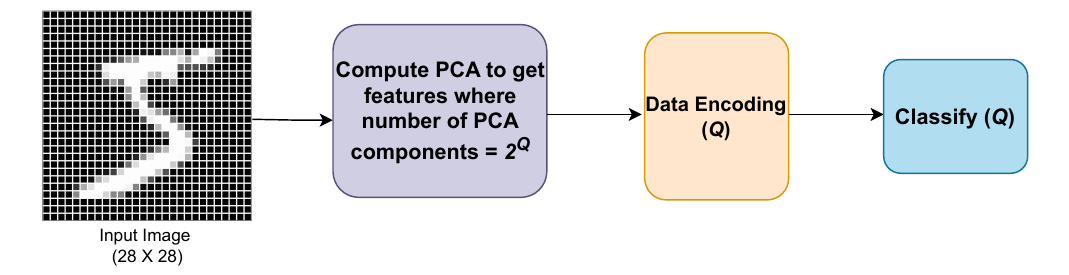}
    \vspace*{-3em}
  \caption{}
  \label{fig:sub5}
\end{subfigure}%

\caption{The dimensionality reduction techniques used to compress the classical data for quantum processing. (a) shows the approach with fully connected layers. (b) shows the approach with maximum pooling. (c) shows the appraoch with average pooling. (d) shows the approach with center cropping, and (e) shows the approach with PCA} 
\label{fig:complete}
\end{figure}

In Sec. \ref{sec:dimen_reduc}, we mention various dimensionality techniques other than using a fully connected (FC) layer that can be applied to the classical data in order to reduce the dimension of the data and prepare it for quantum processing. These techniques are: \textbf{(i)} Center cropping: which includes cropping the classical images at the center with a dimension of $N \times N$, where $N \times N = 2^Q$ such that these $2^Q$ features can be encoded using amplitude encoding. \textbf{(ii)} Principal Component Analysis (PCA): This is a linear dimensionality reduction technique that aims to find a lower-dimensional subspace that captures the maximum variance in the data. PCA projects the high-dimensional data onto a new set of orthogonal axes (principal components) ordered by their variance. By selecting the top principal components, the dimensionality of the data can be reduced while retaining most of the relevant information. \textbf{(iii)} Maximum Pooling: This is a dimensionality reduction technique commonly used in convolutional neural networks. It partitions the input into non-overlapping regions and selects the maximum value from each region, effectively downsampling the input while retaining the most salient features. \textbf{(iv)} Average Pooling: Similar to maximum pooling, average pooling divides the input into non-overlapping regions but instead of selecting the maximum value, it computes the average value within each region. This technique helps to reduce the spatial dimensionality while preserving essential information. A visual representation of the dimensionality reduction techniques along with the original technique applied is shown in Fig. \ref{fig:complete}.

In Table \ref{tab:ablation1}, we present the ablation studies of these techniques using the two student models on all three datasets, compared to the original results. Based on the ablation studies, we make the following observations: \textbf{(i)} Applying dimensionality reduction techniques such as center cropping, PCA, maximum pooling, and average pooling leads to a significant drop in the classification accuracy compared to using the original FC layers. This underperformance is observed across all three datasets (MNIST, FashionMNIST, and CIFAR10) for both the 4-qubit and 8-qubit student quantum models. \textbf{(ii)} The original method of using FC layers outperforms all dimensionality reduction techniques by a substantial margin. This is a testament to the number of learnable parameters in the fully-connected layer, which helps in retaining information better than the other methods. For instance, on the MNIST dataset, the 4-qubit model achieves an accuracy of 77.69\% with FC layers, while the best dimensionality reduction technique (average pooling) achieves only 45.87\%. \textbf{(iii)} Among the dimensionality reduction techniques, average pooling consistently performs better than the other techniques (center cropping, PCA, and maximum pooling) across all datasets and models, suggesting that average pooling is relatively more effective in retaining essential features for quantum processing. \textbf{(iv)} The performance gap between the 4-qubit and 8-qubit models is maintained even when using dimensionality reduction techniques, with the 8-qubit model consistently outperforming the 4-qubit model, indicating that the increased qubit capacity allows for better representation and processing of the input data.

\begin{table}[htbp]
\small
\centering
\caption{Accuracy (\%) of the student quantum models with different dimensionality reduction techniques compared to the original method applied on the three datasets.}
\label{tab:ablation1}
\begin{tabular}{@{\extracolsep{4pt}}lccccc@{}}
\toprule
\multirow{2}{*}{\textbf{Model}} & \multicolumn{5}{c}{\textbf{MNIST}} \\
\cmidrule(r){2-6}
& Center Crop & PCA & Max Pool & Avg Pool & FC layers \\
\midrule
4-qubit & $31.35 \pm 0.80$ & $14.98 \pm 0.60$ & $19.21 \pm 0.70$ & $45.87 \pm 0.90$ & \textbf{77.69 $\pm$ 0.25}\\
8-qubit & $42.72 \pm 0.70$ & $19.30 \pm 0.65$ & $25.36 \pm 0.75$ & $52.50 \pm 0.85$ & \textbf{90.40 $\pm$ 0.15} \\
\midrule
\multicolumn{6}{c}{\textbf{FashionMNIST}} \\
\midrule
4-qubit & $27.91 \pm 0.85$ & $17.01 \pm 0.65$ & $21.11 \pm 0.75$ & $48.29 \pm 0.95$ & \textbf{77.36 $\pm$ 0.30} \\
8-qubit & $32.04 \pm 0.80$ & $18.42 \pm 0.70$ & $18.57 \pm 0.70$ & $50.01 \pm 0.90$ & \textbf{83.44 $\pm$ 0.20}\\
\midrule
\multicolumn{6}{c}{\textbf{CIFAR10}} \\
\midrule
4-qubit & $12.14 \pm 0.90$ & $6.85 \pm 0.70$ & $8.05 \pm 0.80$ & $14.32 \pm 1.00$ & \textbf{25.44 $\pm$ 0.40}\\
8-qubit & $12.33 \pm 0.85$ & $7.99 \pm 0.75$ & $9.83 \pm 0.85$ & $20.73 \pm 0.95$ & \textbf{31.62 $\pm$ 0.35}\\
\bottomrule
\end{tabular}
\end{table}

\subsubsection{Ablation on Data Encoding Techniques}
In Sec. \ref{sec:encode_method}, we use amplitude encoding as the encoding method to map the reduced classical data into quantum states, thus enabling us to run the QNN circuits and perform classification. However, in Sec. \ref{sec:encode_preli}, we discuss two other approaches which are widely used in quantum data encoding, especially for QNNs. In Table \ref{tab:ablation2}, we present the classification results of the two quantum model on the three datasets using the other two encoding techniques, as compared to amplitude encoding. Based on the ablation studies, we make the following observations: \textbf{(i)} The amplitude encoding method outperforms both angle encoding and qubit encoding techniques for all three datasets (MNIST, FashionMNIST, and CIFAR10) and for both the 4-qubit and 8-qubit student quantum models. \textbf{(ii)} Amplitude and qubit encoding consistently performs better than angle encoding, indicating their effectiveness in preserving relevant information for quantum processing. This can be credited to the fact that both amplitude and qubit encoding has the ability to use $2^Q$ features encoded in the quantum states, whereas angle encoding can use only $Q$ features for $Q$ qubits. \textbf{(iii)} The performance gap between the 4-qubit and 8-qubit models is observed across all encoding techniques, with the 8-qubit model achieving higher accuracy than the 4-qubit model, further emphasizing the advantage of increased qubit capacity. \textbf{(iv)} The relative performance differences between the encoding techniques become more pronounced as the dataset complexity increases, with amplitude encoding maintaining a larger advantage over the other techniques on the more challenging CIFAR10 dataset compared to MNIST and FashionMNIST. This indicates that in more complex datasets, there is more room for improvement, and thus, applying better techniques can improve performance significantly.
\begin{table}[!t]
    \small
    \centering
    \caption{Accuracy (\%) of the student quantum models with different data encoding techniques compared to the original amplitude encoding method applied on the three datasets.}
    \label{tab:ablation2}
    \begin{tabular}{@{\extracolsep{4pt}}lccc@{}}
        \toprule
        \multirow{2}{*}{\textbf{Model}} & \multicolumn{3}{c}{\textbf{MNIST}} \\
        \cmidrule(r){2-4}
        & Angle encoding & Qubit encoding & Amplitude encoding \\
        \midrule
        4-qubit & 62.21 $\pm$ 0.17 & 75.04 $\pm$ 0.10 & \textbf{77.69 $\pm$ 0.25}\\
        8-qubit & 78.94 $\pm$ 0.30 & 85.66 $\pm$ 0.08 & \textbf{90.40 $\pm$ 0.15}\\
        \midrule
        \multicolumn{4}{c}{\textbf{FashionMNIST}} \\
        \midrule
        4-qubit & 68.42 $\pm$ 0.13 & 76.22 $\pm$ 0.16 & \textbf{77.36 $\pm$ 0.30}\\
        8-qubit & 71.60 $\pm$ 0.03 & 81.19 $\pm$ 0.31 & \textbf{83.44 $\pm$ 0.20}\\
        \midrule
        \multicolumn{4}{c}{\textbf{CIFAR10}} \\
        \midrule
        4-qubit & 11.90 $\pm$ 0.45& 23.65 $\pm$ 0.02 & \textbf{25.44 $\pm$ 0.40}\\
        8-qubit & 16.14 $\pm$ 0.10& 27.80 $\pm$ 0.34 & \textbf{31.62 $\pm$ 0.35}\\
        \bottomrule
    \end{tabular}
\end{table}

\begin{figure}[!htbp]
    \centering
    \includegraphics[width=1\linewidth]{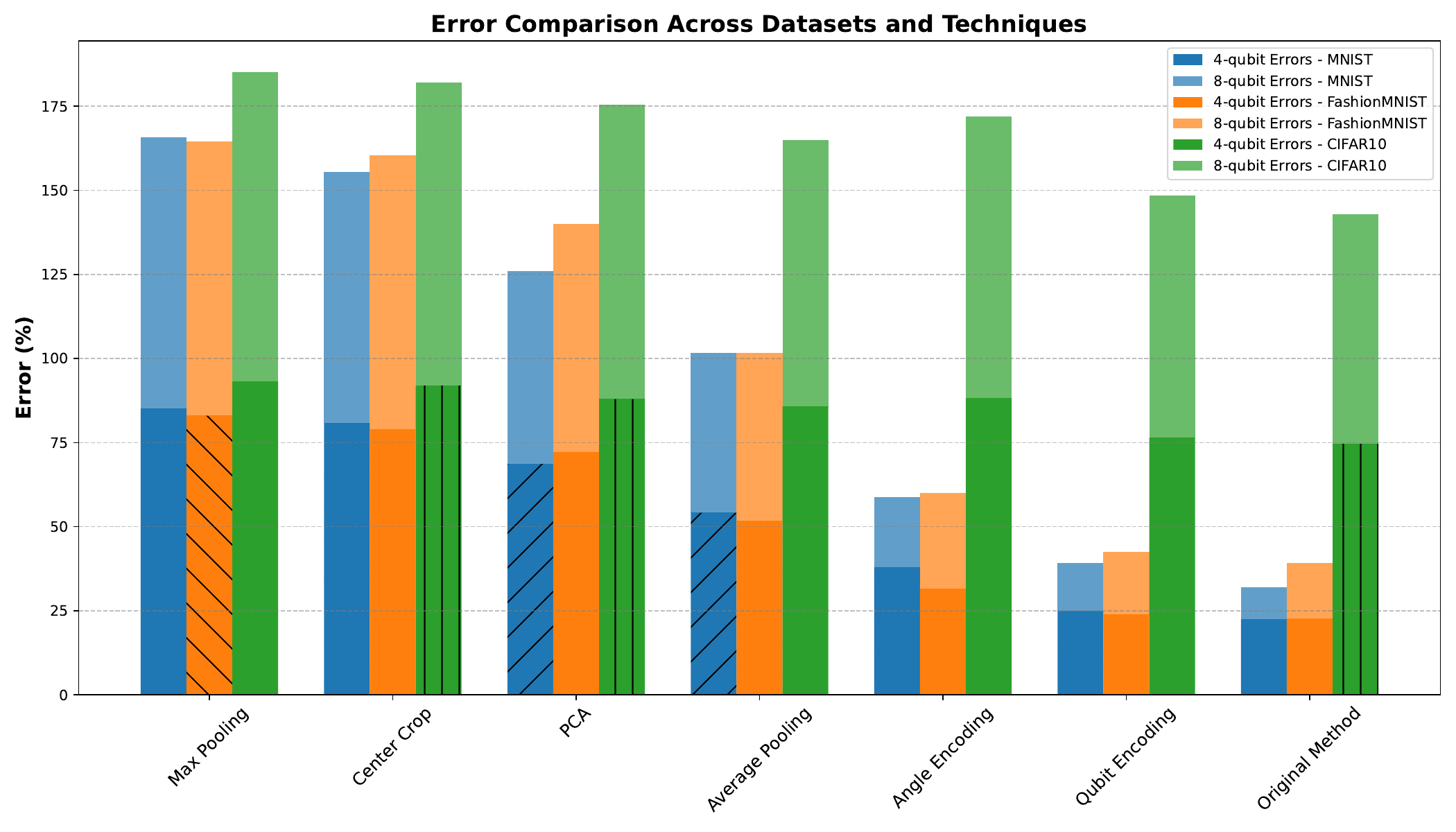}
    \caption{Bar plot visually comparing the performance of the quantum models on all the methods applied along with their ablation studies in ascending order of performance}
    \label{fig:all_results}
\end{figure}

Fig. \ref{fig:all_results} shows an stacked bar plot of the errors percentage of all the methods applied in this study along with their ablation studies, in descending order of performance for the quantum models on the three datasets. The plot reveals that the quantum models perform exceptionally well on the MNIST dataset compared to the FashionMNIST dataset in all the experiments except for maximum pooling. Furthermore, it shows how poorly maximum pooling and center cropping performs compared to the other techniques. Average pooling performs much better than maximum pooling on all three datasets consistently, implying that for quantum circuits, the average presence of features in a patch of image data aids in extracting features better than retaining the most prominent features in the patch. All the encoding methods perform well since fully-connected layers were used to reduce the dimensionality of the features in these methods. However, angle encoding performs much better than qubit encoding, although they both encode similar number of input features in quantum states. Hence, it can be concluded that amplitude encoding extracts features better than qubit encoding across datasets and quantum models for the image classification task.

\section{Discussion}

In our study, we have introduced a novel approach to enhancing the performance of quantum neural networks (QNNs) through knowledge distillation from classical neural networks. A key aspect of our investigation lies in the stark disparity in the number of parameters between the classical (teacher) and quantum (student) models. This section delves into the implications of this disparity, the efficiency of quantum models, and the role of knowledge distillation in bridging the performance gap.

\subsection{Parameter Efficiency and Quantum Advantage}

Table~\ref{tab:parameters} presents a comparison of the number of parameters across different models used in our experiments.

\begin{table}[htb!]
\small
\caption{Comparison of Parameter Counts in Classical and Quantum Models}
\label{tab:parameters}
\centering
\begin{tabular}{lcr}
\toprule
\textbf{Model Type} & \textbf{Model Name} & \textbf{Parameter Count} \\
\midrule
Quantum (Student) & 4-qubit & 3,193 \\
Quantum (Student) & 8-qubit & 6,373 \\
Classical (Teacher) & AlexNet & 94,672,074 \\
Classical (Teacher) & LeNet & 44,426 \\
Classical (Teacher) & ResNet50 & 23,528,522 \\
Classical (Teacher) & DenseNet121 & 6,964,106 \\
Classical (Teacher) & EfficientNetB0 & 4,020,358 \\
\bottomrule
\end{tabular}
\end{table}

The quantum models, despite their significantly lower parameter counts, achieve remarkable performance through the distillation process. For instance, the 4-qubit model, with only 3,193 parameters, benefits from distilled knowledge originating from classical models that possess parameter counts orders of magnitude larger. This efficiency underscores a quantum advantage, where QNNs utilize the principles of quantum mechanics to process information in ways that classical models, even with their extensive parameters, cannot mimic.

\subsection{Implications of Knowledge Distillation}

The process of knowledge distillation, as demonstrated in our results, effectively transfers the ``essence" of the classical models' learning to the quantum models. This not only improves the quantum models' accuracy but also illustrates a crucial point: quantum models can achieve competitive performance levels with a fraction of the parameters used by classical models. This efficiency is pivotal for applications with stringent resource constraints and points towards a future where hybrid classical-quantum systems could become commonplace.

Furthermore, the scalability observed with the increase in qubits from 4 to 8 suggests that as quantum technologies advance, we can expect even more significant performance enhancements. This scalability, combined with the inherent efficiency of quantum models, sets the stage for a broader adoption of quantum computing in machine learning tasks.

\subsection{Future Work}
Looking ahead, our research lays the groundwork for several promising trajectories:

\textbf{Quantum-inspired Distillation Techniques:} While traditional knowledge distillation techniques have been effective, there is a vast unexplored territory in the realm of quantum-inspired distillation strategies. Tailoring these methods to the unique characteristics of quantum neural networks could further optimize the knowledge transfer process.

\textbf{Co-Evolutionary Quantum-Classical Architectures:} This study concentrated on a unidirectional transfer of knowledge from classical to quantum models. An intriguing future direction would be to investigate co-evolutionary frameworks where quantum and classical networks mutually enhance each other's learning, potentially leading to a quantum leap in the capabilities of quantum machine learning systems.

\section{Conclusion}
\label{sec:con}
In this paper, we have explored a new approach for bridging the gap between classical and quantum neural networks through knowledge distillation. Our approach has demonstrated that classical networks can effectively serve as mentors to their quantum counterparts, significantly boosting the latter's performance on benchmark image classification datasets such as MNIST, FashionMNIST, and CIFAR-10. Remarkably, these enhancements were realized even though the quantum student networks utilized far fewer parameters than the classical teacher networks.

The implications of our findings are profound for the evolving field of quantum machine learning. The parameter efficiency of quantum neural networks, when augmented by the knowledge distilled from classical networks, heralds a new era of possibilities for quantum computing in scenarios where resources are at a premium. Moreover, the observed scalability with the addition of more qubits hints at the untapped potential that could be unleashed as quantum computing hardware continues to evolve. Thus, our work serves as a stepping stone towards realizing the full potential of quantum neural networks and sets the stage for a future where quantum and classical machine learning coalesce to solve complex computational challenges.

\section*{Statements and Declarations}

\appendix

\subsection*{Conflict of interest}
The authors declare that they have no known competing financial interests or personal relationships that could have appeared to influence the work reported in this paper.

\subsection*{Funding}
This research received no specific grant from funding
agencies in the public, commercial, or not-for-profit sectors.

\subsection*{Data availability}
The datasets used in this study are publicly available and can be downloaded or accessed through machine learning frameworks like PyTorch or Tensorflow.


\subsection*{Acknowledgements}
The authors acknowledge the support of NSU CTRGC grant 2023-24 approved by the NSU authority.


\end{document}